%% file: ee2013.tex
\documentclass[preprint2]{emulateapj}

\usepackage{amsmath}
\usepackage{natbib}
\usepackage{graphicx}
\usepackage{epsf}
\usepackage{subfigure}
\usepackage{color}
\usepackage{threeparttable}
\usepackage{comment}
\usepackage{epsfig}
\usepackage{xspace}
\usepackage{hyperref}
\usepackage[usenames,dvipsnames,svgnames]{xcolor}
\usepackage{latexsym}
\DeclareGraphicsExtensions{.jpg,.pdf,.png,.eps,.ps}

 \newcommand{\LCDM}{\mbox{$\Lambda$CDM}\xspace}

 \newcommand{\ltsima}{$\; \buildrel < \over \sim \;$}
 \newcommand{\ltsim}{\lower.5ex\hbox{\ltsima}}

 \newcommand{\sqdeg}{\ensuremath{\mathrm{deg}^2}}

 \newcommand{\neff}{\ensuremath{N_\mathrm{eff}}\xspace}

 \newcommand{\wmap}{\textit{WMAP}\xspace}
 
 \newcommand{\planck}{\textit{Planck}}

 \newcommand{\degs}{\ensuremath{\mathrm{deg}^2}}
 
 \newcommand{\simleq}{{\raise.0ex\hbox{$\mathchar"013C$}\mkern-14mu \lower1.2ex\hbox{$\mathchar"0218$}}}
 \newcommand{\simgeq}{{\raise.0ex\hbox{$\mathchar"013E$}\mkern-14mu \lower1.2ex\hbox{$\mathchar"0218$}}}

 \def\microKsq{\mu{\mbox{K}}^2}
 \def\microK{\mu{\mbox{K}}}

 \def\mb{\mathbf}

\begin{document}
\title{MEASUREMENTS OF E-MODE POLARIZATION AND TEMPERATURE-E-MODE CORRELATION IN THE COSMIC MICROWAVE BACKGROUND FROM 100 SQUARE DEGREES OF SPTPOL DATA}

\input ee2013_authorlist_v3.tex

\begin{abstract}

We present measurements of $E$-mode polarization 
and temperature-$E$-mode correlation in the cosmic microwave background (CMB) 
using data from the first season of observations with SPTpol, the polarization-sensitive 
receiver currently installed on the South Pole Telescope (SPT). 
The observations used in this work cover 100~\sqdeg\ of sky with arcminute resolution at $150\,$GHz.
We report the $E$-mode angular auto-power spectrum ($EE$)
and the temperature-$E$-mode angular cross-power spectrum ($TE$)
over the multipole range $500 < \ell \leq5000$.
These power spectra improve on previous measurements in the
high-$\ell$ (small-scale) regime.
We fit the combination of the SPTpol power spectra, data from \planck\, 
and previous SPT measurements with a six-parameter \LCDM cosmological model.
We find that the best-fit parameters are consistent with previous results.
The improvement in high-$\ell$ sensitivity over previous measurements
leads to a significant improvement in the limit on polarized
point-source power:
after masking sources brighter than 50\,mJy in unpolarized flux at 150\,GHz, we find
a 95\% confidence upper limit on unclustered point-source power in the $EE$ spectrum of 
$D_\ell = \ell (\ell+1) C_\ell / 2 \pi < 0.40 \ \microKsq$
at $\ell=3000$, indicating that future $EE$ measurements will not be limited by
power from unclustered point sources in the multipole range $\ell < 3600$, and 
possibly much higher in $\ell.$


\end{abstract}


\maketitle

\section{Introduction}
\setcounter{footnote}{0}

Measurements of the cosmic microwave background (CMB) are a cornerstone of our understanding of cosmology.  
The angular power spectrum of the CMB temperature anisotropy has now been measured to high precision at 
scales of tens of degrees 
down to arcminutes with, e.g., the \textit{Wilkinson Microwave Anisotropy Probe} (\wmap) and
\planck\ satellites \citep{bennett13,planck13-15} and the ground-based
Atacama Cosmology Telescope (ACT) and South Pole Telescope (SPT) experiments
\citep{das14,story13}.  These measurements yield tight constraints on cosmology \citep[e.g.][]{hinshaw13,planck13-16,sievers13,hou14}.

The CMB is also partially polarized. The largest contribution to the polarization was
imprinted during the epoch of recombination, when local
quadrupole intensity fluctuations, incident on free electrons, created linear polarization via Thomson scattering
\citep[e.g.,][]{hu97d}.  The observed linear polarization pattern in the CMB 
can be decomposed into even-parity ($E$-mode) and odd-parity ($B$-mode) components, 
with a key feature being that scalar (density) perturbations during recombination generate $E$-mode polarization patterns only.  By contrast, the local quadrupoles induced by gravitational waves generally produce both $E$ and $B$ modes.  Gravitational lensing will also generate $B$ modes by distorting the scalar-induced $E$-mode pattern and converting a small fraction into $B$. The total $B$-mode contribution from
these two mechanisms is measured/limited to be at least an order of magnitude smaller
in amplitude (two orders of magnitude smaller in power) than the $E$ modes 
\citep{hanson2013,polarbear2014b,bicep2a}.

Measurements of the $E$-mode polarization of the CMB are complementary to intensity measurements in a number of ways. Particularly relevant to this paper is an expected advantage in CMB signal-to-foreground contamination ratio at small angular scales, 
due to the low level of expected polarization in emission from extragalactic point sources.
We expect that, as experiments reach lower noise levels, polarization measurements
will extend to angular scales that are smaller than those limited by
point source contamination in temperature.

The $E$-mode polarization of the CMB and the temperature-$E$-mode correlation were first detected 
in data from the Degree Angular Scale Interferometer (DASI; \citep{kovac02}).
Many experiments have since reported measurements of the angular auto-power spectrum of 
the $E$ modes (the $EE$ spectrum) 
and the angular cross-power spectrum between the temperature and the $E$ modes (the $TE$ spectrum).
The best measurements of the $EE$ and $TE$ spectra to date  
come from \wmap\ \citep{bennett13} and BICEP2 \citep{bicep2a}
on scales of a degree and larger
and from ACTpol on sub-degree scales \citep{naess14}.
At the precision of these current measurements, the $EE$ and $TE$ power
spectra are consistent with the predictions from cosmological models
that are highly constrained by the temperature power spectrum ($TT$).

In this work, we present estimates of the $EE$ and $TE$ angular power spectra 
using 100~\sqdeg\ of data from the first season of observations with SPTpol, 
a polarization-sensitive receiver installed on the SPT. We report measurements
in the multipole range $500 < \ell \le 5000$, with high signal-to-noise on the 
primary anisotropy power
spectra out to $\ell \simeq 3000$.

This paper is structured as follows: 
Section 2 describes the instrument and the observations. 
Section 3 describes the low-level data processing and calibration. 
Section 4 outlines the steps used to transform time-ordered data into maps and power spectra. 
Section 5 describes our calibration procedures. 
Section 6 discusses the calculation of the bandpower covariance matrix. 
Section 7 describes  the results of jackknife and other systematic tests.  
Section 8 presents our $EE$ and $TE$ spectra. 
Section 9 describes cosmological constraints. 
Finally, Section 10 states our conclusions.

\section{The SPTpol Instrument}
\label{instrument}

SPTpol is a polarization-sensitive receiver installed in early 2012 on the SPT \citep{carlstrom11}.  The receiver was designed to make precision measurements 
of the polarization of the CMB over a wide range of angular scales.  It replaced the original instrument installed on the telescope, the SPT-SZ receiver, which was sensitive only to intensity fluctuations.

The SPT is a 10-meter off-axis Gregorian design with a one-square-degree
field of view and arcminute resolution at 150\,GHz \citep{padin08}.
The field of view and resolution make
the SPT an ideal telescope for surveying large areas and
measuring the anisotropy of the CMB from degree scales
out to arcminute scales. The Gregorian focus allows for an
optical design with nearly zero cross-polarization at the center of 
the focal plane \citep{mizuguchi78,dragone82}.
The optical design of the SPT and 
the SPTpol receiver are described in detail in \citet{padin08} and \citet{george12}, respectively.

The SPTpol focal plane contains 1536 polarization-sensitive
transition edge sensor (TES) bolometers, with
1176 detectors at 150\,GHz and 360 detectors at 95\,GHz.
The detectors are operated at $\sim500$\,mK and are read out with a digital frequency-domain multiplexing readout \citep{dobbs08,dobbs12b,dehaan12}.
The detectors in the two bands were designed and fabricated 
independently and are described in detail in \citet{henning12} (150\,GHz) and 
\citet{sayre12} (90\,GHz).  This work focuses on data from the 150\,GHz array.

The 150\,GHz array is composed of seven detector modules,
each containing 84 pixels. 
The modules were fabricated at the National Institute for Standards and Technology.
Each consists of a  2.3~inch-wide hexagonal detector array behind a monolithic feedhorn array.  Incoming 
power is coupled through the feedhorns to an
orthomode transducer (OMT),
which splits the light into two orthogonal polarization states.  
The signal from each polarization state is coupled via microstrip to a thermally isolated detector island.  
Changes in island temperature are read out by an aluminum manganese TES
with a transition temperature of $\sim 500$\,mK
connected to an array of superconducting quantum interference device (SQUID) amplifiers.  
For additional details on the SPTpol instrument design,
characterization, and operation also see
\citet{george12,austermann12,story12}.

\section{Observations and Data Reduction}
\label{sec:data}
\subsection{Observing Strategy}
\label{sec:obs}

The first year of observations with the SPTpol camera in 2012 
focused on a 100~\sqdeg\ patch of sky centered at right ascension 23h30m and declination $-55$ degrees.
We refer to this field as the SPTpol ``100d'' field to distinguish it from the full 500~\sqdeg\ survey field 
which we began observing in 2013.  All observations of the SPTpol 100d field were made using an azimuthal ``lead-trail'' observing strategy.
In this observing strategy the field is split into two equal halves in right ascension, a ``lead'' half-field and a ``trail'' half-field.
The lead half-field is observed first, followed immediately by a trail half-field observation, with the timing adjusted such 
that the lead and trail observations cover the same azimuth range. Each half-field is observed
by scanning the telescope in azimuth back and forth across the field and then stepping up in elevation.
We refer to one pass of the telescope, either from left to right or from right to left across the field, as a ``scan,''
and to a half-hour set of scans that cover an entire half-field as an ``observation.''
In the 2012 observing season, the full field was observed roughly 2500 times. The azimuthal
scanning speed used in all observations was 0.48 degrees per second, corresponding to 0.28 
degrees per second on the sky at the mean elevation of the field.

This observing strategy enables removal of ground pickup via the differencing of pairs of lead and trail observations \citep[e.g.,][]{pryke09}.
Tests for ground pickup in the 2012 SPTpol data (including the jackknife null tests described in 
Section~\ref{sec:nulltests}) show no sign of significant ground contamination, so we do not 
use a field-differencing analysis in this work. There is a small overlap region between the lead 
and trail half-fields, leading to a region of deeper coverage (lower noise) in the middle of the full
coadded field.  To simplify the analysis, we ignore this and use a uniform weight over the field in
the power spectrum calculation described in Section \ref{sec:ps}.

\subsection{Map-Making: Time-Ordered Data to Maps}
\label{sec:mapmaking}
The treatment of the time-ordered data (TOD) and the map-making process
is similar to that in analyses of data from SPT-SZ \citep[e.g.,][]{lueker10}, with the added complexity of
calculating all three Stokes parameters $I$, $Q$, and $U$
in each map pixel.  We will refer to the Stokes $I$
parameter as $T$ (temperature) for the rest of this paper and express $Q$
and $U$ in temperature units.  The procedure for polarized map-making follows the methodology described in \citet{couchot99}, \citet{montroy03}, \citet{jones07}, and \citet{chiang10}.

The TOD are bandpass-filtered by applying a low-pass filter 
and subtracting a fourth-order polynomial from each scan (effectively a high-pass filter).  The cutoff frequencies for these two filters correspond to angular frequencies in the scan
direction of $\ell \simeq 10000$ and $\ell \simeq 150$ at the mean elevation of the field.\footnote{Throughout
this work, we use the flat-sky approximation to equate multipole number $\pmb{\ell}$ with $2 \pi |\mathbf{u}|$, where 
$\mathbf{u}$ is the Fourier conjugate of Cartesian angle on an asymptotically flat patch of sky.}
The data from each detector are calibrated relative to each other
using the method described in Section \ref{sec:relcal}, 
and then combined into $T$, $Q$, and $U$ maps
using the pointing information, polarization angle, and weight for each detector.
Weights are calculated for each observation based on detector polarization efficiency and 
noise power between 1~Hz and 3~Hz, 
which corresponds to the angular scales of the signals of interest.
The noise power is calculated by taking the difference between left-going and right-going scans.
A 3-by-3 matrix representing the $T$, $Q$, and $U$ weights and the correlations between the three
measurements is created for each map pixel using this same information. 

We make maps using the oblique Lambert azimuthal equal-area projection with 0.5-arcminute pixels.
This sky projection introduces small angle distortions, which we account for by rotating the 
$Q$ and $U$ Stokes components across the map to maintain a consistent angular coordinate system in this projection.
We combine $Q$ and $U$ maps in Fourier space to create $E$ maps using the standard convention \citep{zaldarriaga01}
\begin{equation}
\label{eqn:def_E}
E_{\pmb{\ell}} = Q_{\pmb{\ell}} \cos{2\phi_\ell} + U_{\pmb{\ell}} \sin{2\phi_\ell}, 
\end{equation}
where $\phi_\ell$ is the azimuthal angle of $\pmb{\ell}$. 

\subsubsection{Cuts and Flagged Data}
\label{sec:cuts}
Data are included in the analysis if they pass several quality checks.
We perform checks at several different levels:
individual-bolometer TOD of a single scan,
individual-bolometer data over full observations,
and full observations which include all bolometers' TOD that have passed cuts.

Data from individual bolometers are flagged on a per-scan basis  
based on the presence of
various types of 
discontinuities in the data, and on the noise in the TOD.  Scans are flagged for a particular bolometer if either a sharp spike (presumably caused by 
a cosmic ray) or a sharp change in DC level (attributed generally to changes in SQUID bias point)
is detected above a given significance threshold.  After removing flagged scans, 
the root mean square (rms) of each bolometer's data on each scan is calculated.
The median rms for bolometers within the same module for an entire observation is calculated, and
scans with an rms greater than 3.5 times this median or less than 0.25 times this median are flagged.
This last flagging step is done after the polynomial subtraction described in the previous section.
Data from scans that are flagged for any of these reasons are not included in maps, removing 5 percent of the data.

Data from individual bolometers are
flagged in full observations based on their response and noise properties.  
Detectors with low signal-to-noise response to either of two regularly performed calibration
observations (two-degree elevation dips and observations of an internal blackbody source) are flagged for the entire subsequent CMB field observation.  
We similarly flag bolometers with abnormally high or low noise in the $1-3$~Hz frequency band. 

Single-observation maps (using data from all bolometers and scans not already flagged)
are flagged based on the values of the following (often coincident) criteria:
rms noise in the map, median pixel weight, 
the product of median weight and map noise squared, 
and the sum of the weights over the full map.
For these cuts, we remove outliers both above and below the median value for each field.
We do not use observations that are flagged by one or more of these cuts.
The cut levels are set to exclude data where the values of the above 
criteria are in the tails of the distributions.  
Such outliers typically correspond to observations in which a significant fraction of the array was improperly voltage-biased (often caused by bad or variable weather) and would be equally identified by any of the above cuts.  Due to very conservative cut values (described above) for this first analysis, of the roughly 5000 lead and trail observations of the SPTpol 100d during the 2012 season, 3416 are used in this analysis.

\subsubsection{Coadded Maps}
\label{sec:coaddmaps}
For this analysis we choose to combine multiple observations of the 
CMB field so that the fundamental unit for the power spectrum analysis is a coadd of 28 observations, 
which we refer to as a ``map bundle''.
We combine individual-observation 
maps into bundles such that every bundle has nearly uniform coverage on the field.  
After cuts, there are 122 map bundles for the 2012 data.  

\begin{figure*}
\begin{center}
\includegraphics[width=0.9\textwidth]{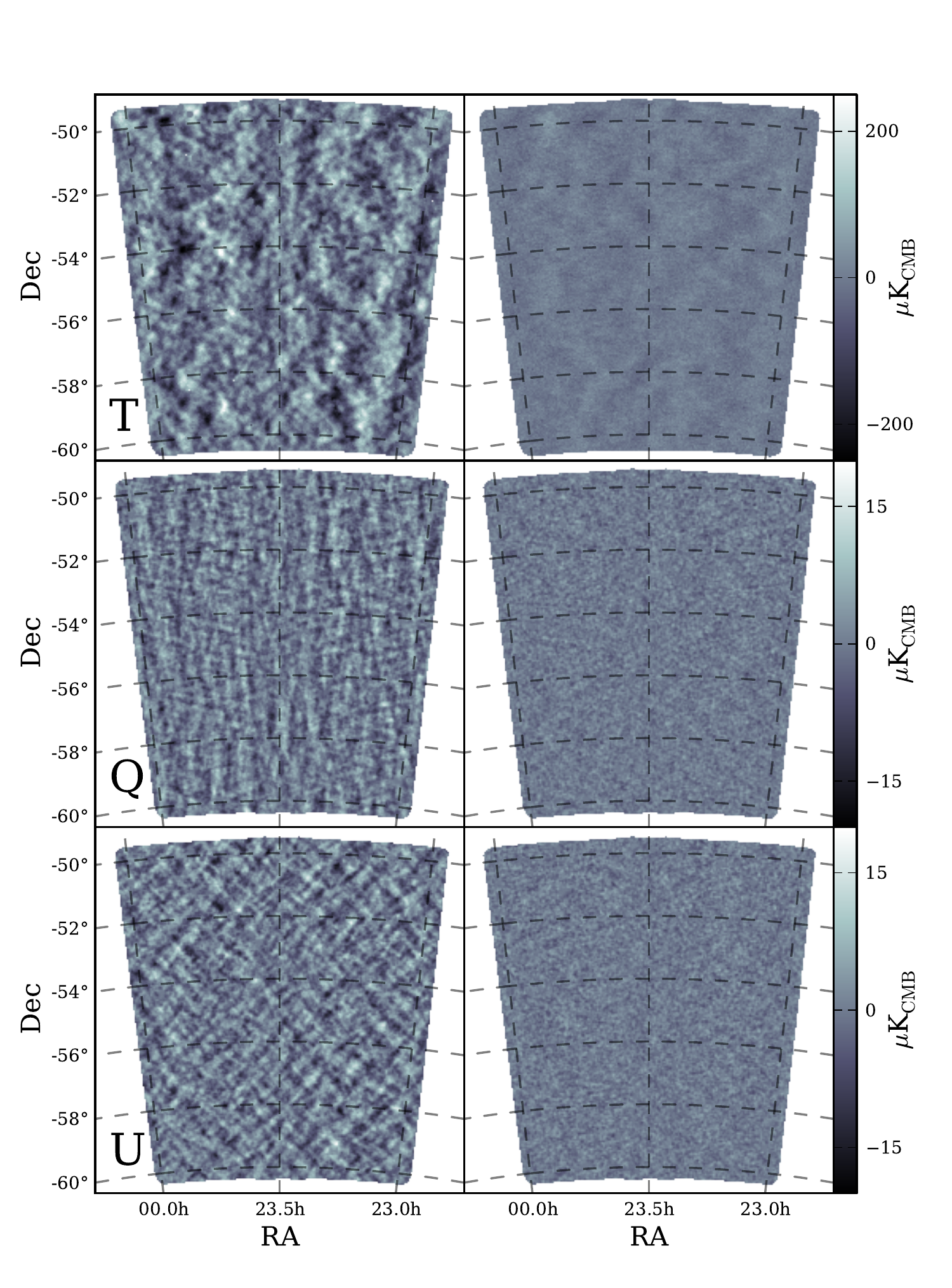}
\caption{SPTpol 2012 100d signal (Left) and noise (Right) maps.  
The noise maps are obtained by subtracting data from the first half of the season from data from the second half and dividing by two. 
(Top) $T$ maps, $\pm250$ $\mu$K$_{\mathrm{CMB}}$.  (Middle) Stokes $Q$ maps, $\pm20$ $\mu$K$_{\mathrm{CMB}}$.  
(Bottom) Stokes $U$ maps, $\pm20$ $\mu$K$_{\mathrm{CMB}}$.  
The clear vertical stripes in $Q$ and $\pm45^{\circ}$ stripes in $U$ are indicative of high signal-to-noise $E$ modes.  Lack of horizontal stripes in $Q$ is the result of polynomial subtraction along the scan direction, which is mostly horizontal in this projection.
$Q$ and $U$ maps have been smoothed by a 4.0 arcminute FWHM Gaussian.}
\label{fig:coadds_tqu}
\end{center}
\end{figure*}

\begin{figure*}
\begin{center}
\includegraphics[width=0.75\textwidth]{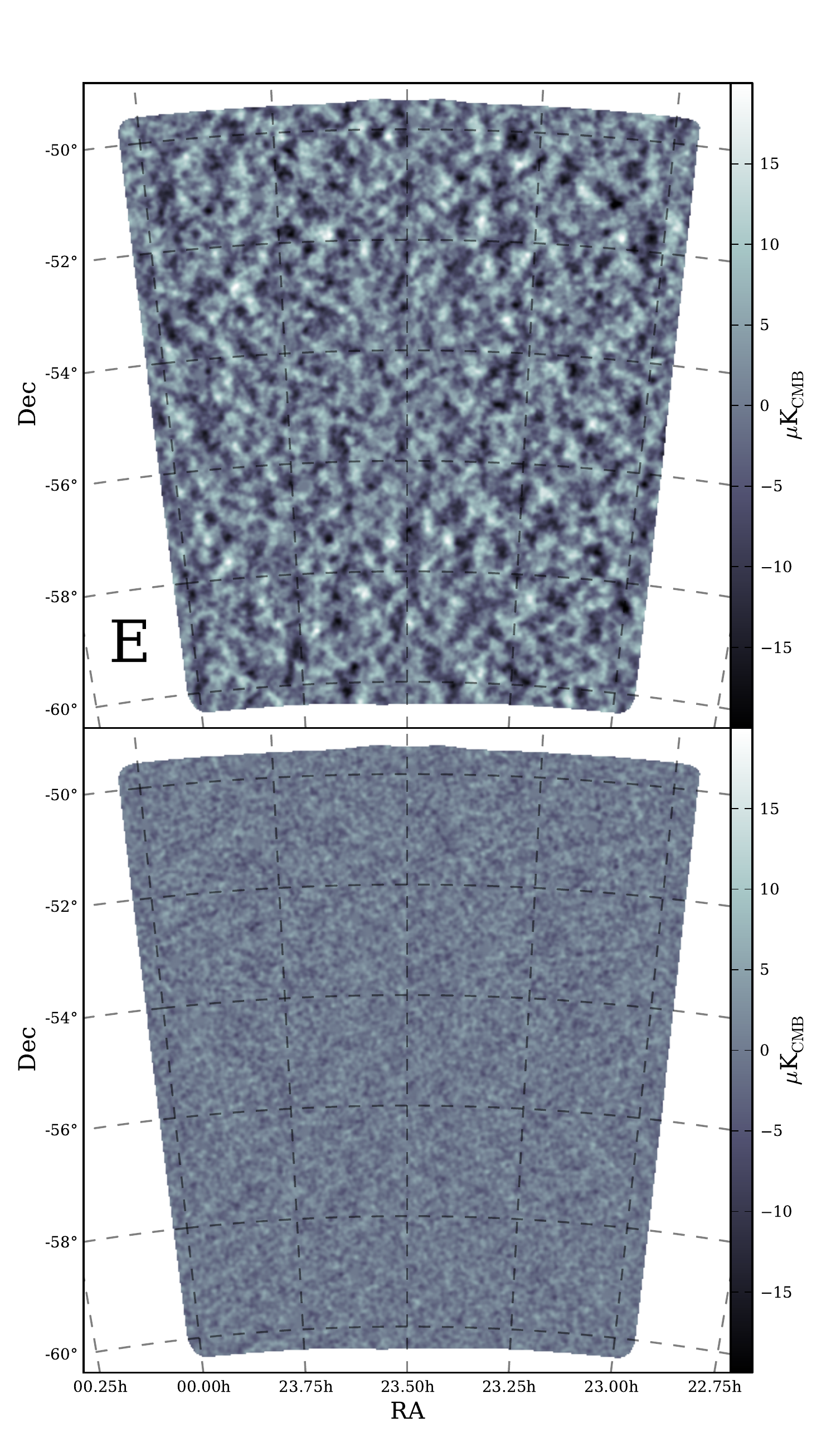}
\vspace{-0.25in}
\caption{SPTpol 2012 100d $E$-mode signal (Top) and noise (Bottom) maps.  
These maps are created from the $Q$ and $U$ signal and noise maps shown 
in Figure \ref{fig:coadds_tqu}, using Equation \ref{eqn:def_E}.
The color scale is $\pm20$ $\mu$K$_{\mathrm{CMB}}$, and both maps have been smoothed by a 4.0 arcminute FWHM Gaussian. }
\label{fig:coadds_e}
\end{center}
\end{figure*}

The power spectrum analysis described in the next section is performed on the map bundles; we  
make full-season coadded maps for display purposes.
Full-season coadds of temperature $T$ and Stokes $Q$ and $U$ maps are shown in the left column of 
Figure \ref{fig:coadds_tqu}.  In the right column, we show maps of one half-season of data subtracted from 
the other half-season (divided by two) to demonstrate the noise level expected in the maps on the left.
Analogous signal and noise maps for $E$-mode polarization are shown in Figure \ref{fig:coadds_e}.  All maps are displayed in units of $\mu$K$_{\mathrm{CMB}}$, which are the equivalent fluctuations of a 2.73\,K blackbody that would produce the measured deviations in intensity.

\section{Power Spectrum}
\label{sec:ps}
In this section, we describe the analysis used to calculate the $EE$ and $TE$ power spectra
from the maps described in the previous section.

\begin{figure}
\begin{center}
\vspace{-0.3in}
\includegraphics[width=0.5\textwidth]{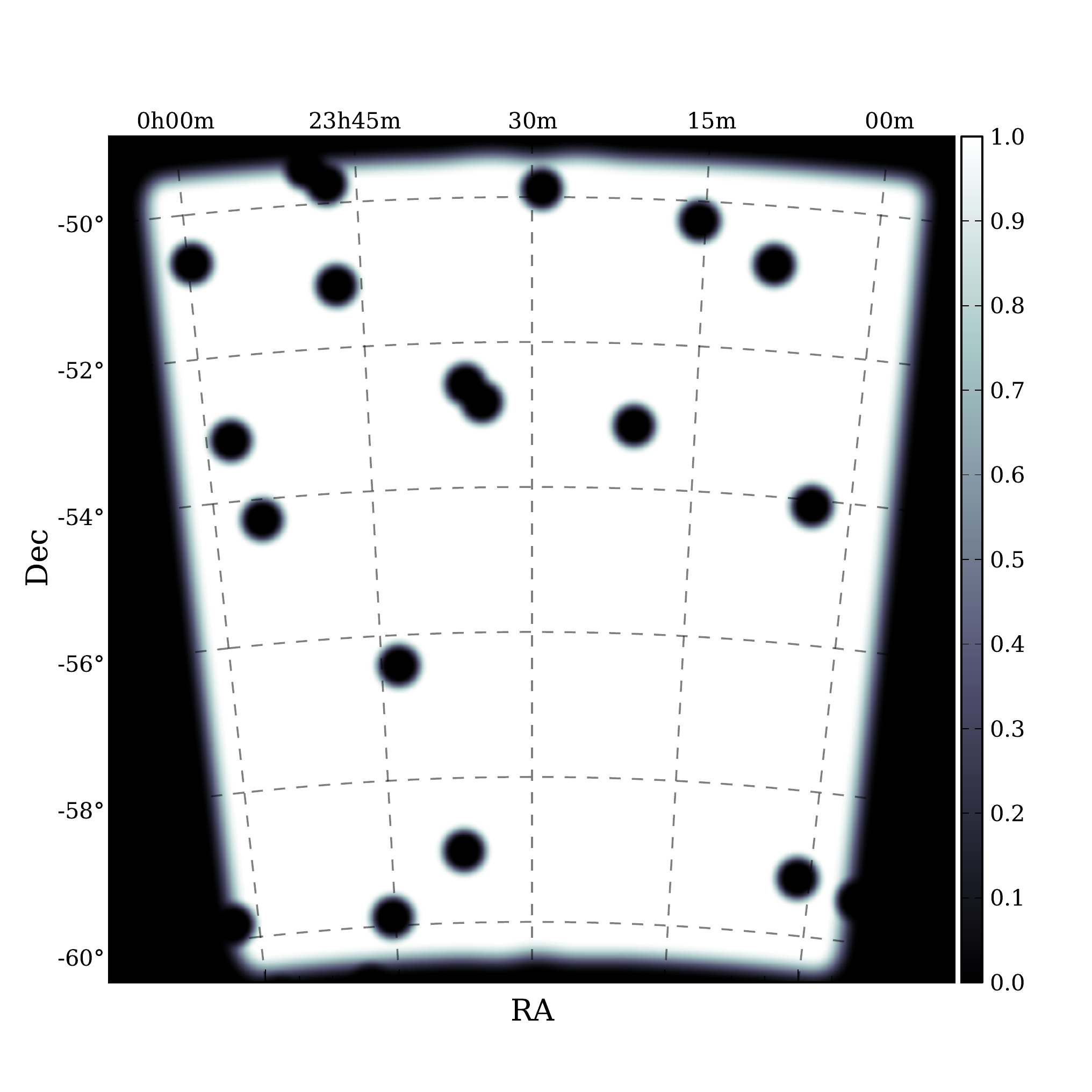}
\vspace{-0.35in}
\end{center}
\caption{The source and apodization mask used to mask the SPTpol 100d maps before calculating power spectra.  It is generated using the procedure described in Section \ref{sec:window}.  All point sources in the field with unpolarized flux $>$ 50\,mJy at 150\,GHz have been masked.
}
\label{fig:mask}
\end{figure}

\subsection{Source and Apodization Mask}
\label{sec:window}
The temperature and polarization maps are multiplied in real space by a two-dimensional array prior to computing the Fourier transform of these maps.  
We refer to this array as the source and apodization mask.  
This mask downweights high-noise regions near the border of the map and effectively removes the flux from a number of bright point sources located in this field.
The list of masked bright point sources consists of all point sources in the field with 150 GHz flux $>$ 50 mJy as measured by SPT-SZ (see \citealt{story13}).
There are 19 sources that are masked using this flux limit in our survey area.
Each point source is masked with a $5'$-radius disk. 
The disk is tapered to zero using a $15'$ cosine taper.
We then smooth the mask with an edge taper of $30'$.
The final mask, which is the product of the apodization and point source masks, is shown in Figure \ref{fig:mask}.

\subsection{Cross-Spectra}
\label{sec:crossspectra}

A pseudo-$C_\ell$ method is used to estimate the binned power spectrum following the MASTER method described in \citet{hivon02}.
To avoid noise bias we use a cross-spectrum analysis \citep{polenta05,tristram05} in which we 
take the cross-spectrum of the subsets of data (bundles) described in the previous section.
These bundles have independent realizations of the atmospheric and detector noise.
We follow the procedure developed in \citet{lueker10}
and used in subsequent SPT power spectrum analyses.

To calculate the power spectrum, we cross-correlate the bundled CMB maps in Fourier space.
The relatively small size of the maps (10 degrees on a side) and the fact that 
our analysis only extends to a minimum multipole number of $\ell=500$ means
that we can use the flat-sky approximation and substitute two-dimensional Fourier transforms 
for spherical harmonic transforms in calculating $C_\ell$.
Each map is multiplied by the source and apodization mask, zero-padded to 1728 by 1728 pixels, 
and the Fourier transform of the map, $\tilde{m}^{A}$, is calculated. 
The resulting Fourier-space maps have pixels of size $\delta_{\ell}=25$ on a side.
We calculate the average cross-spectrum between the maps of two observations $A$ and $B$ within an $\ell$-bin $b$:
\begin{equation}
\label{eqn:ddef}
 \widehat{D}^{AB}_b\equiv \left< \frac{\ell(\ell+1)}{2\pi}Re[\tilde{m}^{A}_{\pmb{\ell}}\tilde{m}^{B*}_{\pmb{\ell}}] \right>_{\ell \in b},
\end{equation}
where $\pmb{\ell}$ is a vector in two-dimensional $\ell$-space and $\ell = |\pmb{\ell}|$. 
There are 122 bundle maps in the input set (see Section \ref{sec:coaddmaps}), resulting in $\sim 10000$ cross-spectra.
We average all cross-spectra $\widehat{D}^{AB}_b$ with $A \neq B$ to calculate a binned power spectrum $\widehat{D}_b$. We refer to these one-dimensional binned power spectrum measurements as ``bandpowers."

\subsection{Unbiased Spectra}
\label{sec:UnbiasedSpectra}

The bandpowers $\widehat{D}_b$ are a biased estimate of the true binned sky power, $D_{b^\prime}$, due to effects such as TOD filtering, beam smoothing, finite sky coverage, and mode-mode mixing from the source and apodization mask.
The biased and unbiased estimates are related by

\begin{equation}
\widehat{D}_{b} \equiv K_{bb^\prime} D_{b^\prime} \, ,
\end{equation}
where the $K$ matrix accounts for the effects of the instrumental beam ($B_{\ell}$), TOD filtering
($F_{\ell}$), and applying the apodization mask ($\textbf{W}$).
$K$ can be expanded as
\begin{equation}
\label{eqn:kdef}
K_{bb^\prime}=P_{b\ell}\left(M_{\ell\ell^\prime}[\textbf{W}]\,F_{\ell^\prime}B^{2}_{\ell^\prime}\right)Q_{\ell^\prime b^\prime}.
\end{equation}
$P_{b\ell}$ is the binning operator and $Q_{\ell^\prime b^\prime}$ its reciprocal \citep{hivon02}.
The mode coupling kernel $M_{\ell\ell^\prime}[\textbf{W}]$ accounts for the mixing of power between bins due to the 
real-space mask applied to the data before Fourier transforming.
The mode coupling kernel is calculated analytically, as described in the Appendix.

\subsection{Bandpower Window Functions}
\label{sec:windowfunc}
Bandpower window functions are necessary to compare the measured bandpowers to a theoretical power spectrum.
The window functions, ${\mathcal W}^b_\ell / \ell$, are defined through the relation
\begin{equation}
C_b = ({\mathcal W}^b_\ell / \ell) C_\ell.
\end{equation}
Following the formalism described in Section~\ref{sec:UnbiasedSpectra}, we can write this as
\begin{equation}
C_b = (K^{-1})_{bb'} P_{b' \ell'} M_{\ell' \ell} F_\ell B_\ell^2 C_\ell,
\end{equation}
which implies that
\begin{equation}
  {\mathcal W}^b_\ell / \ell = (K^{-1})_{bb'} P_{b' \ell'} M_{\ell' \ell} F_\ell B_\ell^2.
\end{equation}

\subsubsection{Simulations and the Transfer Function}
\label{sec:transfer}
We compute the filter transfer function, $F_{\ell}$, with  simulated observations using the iterative method described in \citet{hivon02}.
We start by generating 782 realizations of the CMB sky.
The input cosmology to the simulations is computed using \textsc{CAMB}, a Boltzmann code for calculating CMB power spectra \citep{lewis00}.  
Input parameter values to \textsc{CAMB} are from the \textsc{Planck+lensing+WP+highL} best-fit model in Table 5 of \cite{planck13-16}. 
These maps are then ``observed'' using the actual detector pointing
information to create simulated, noise-free timestreams for each
bolometer.
The simulated timestreams include the same pointing information,
weights, and data cuts as the real data for each observation.
For each scan in each observation, we filter the simulated data in the same way the real data are filtered.
We then make a map for each observation from the filtered data.
The final output of the simulation is 782 realizations for each of the 122 map bundles.

For each of the 782 realizations of the CMB, we combine all of the simulated observations over the full season of data and compute the power spectrum of this coadded map.  
The resulting transfer function $F_\ell$ is the output power spectrum divided by the input power spectrum, after
correcting the output spectrum for mode-mode coupling (see the Appendix).  The $TE$ transfer function is particularly sensitive to locations of zero-crossings in the $TE$ power spectrum, which can cause sharp features in the transfer function unrelated to the effects of filtering on the data.  To avoid these spurious features, we instead define the $TE$ transfer function as the geometric mean of the $TT$ and $EE$ transfer functions.  In the multipole range considered in this analysis, both the $EE$ and $TE$ transfer functions are always greater than 0.7.

\subsubsection{Beam Functions}
\label{sec:beams}
A measurement of the SPT beam -- the optical response as a function of angle -- is needed to calibrate the angular power spectrum as a function of multipole.
The beam response, $B_{\ell}$, for the 2012 instrument is measured from the combination of dedicated observations of Mars and the brightest point sources in the CMB field.

The effective beam for observations of the 100d CMB field is the convolution of the instantaneous response function of the 
system and the effect of random, uncorrected pointing variations between observations of the field. 
The instantaneous beam is measured using eight dedicated observations of Mars from the fall of 2012.
These observations are short enough that any pointing variation over an observation is negligible, and the results from 
the eight observations are registered to one another with high precision before they are combined.
This instantaneous beam is compared to the beam measured from bright 
point sources in the coadded CMB field to estimate the additional beam width 
contributed by pointing variation, or jitter.
In the 2012 data, we measure $12^{\prime \prime}$ rms jitter.  (Recent updates to our pointing reconstruction should reduce the jitter in future analyses).  The convolution of the
Mars-derived beam with a Gaussian of this width provides a good fit
to the profile derived from sources in the field.

We use the jitter-convolved Mars beam map to calculate the beam function $B_{\ell}$, the azimuthally averaged Fourier transform of the beam map.
The uncertainty on the beam measurements is calculated from the standard deviation between individual Mars observations.  Fractional beam uncertainties are less than 1\% over the multipole range reported here. The FWHM of the 150~GHz jitter-convolved beam is 1.18 arcminutes.  

Electrical cross-talk between detectors affects the measured beam and can potentially result in a 
different effective beam for temperature and polarization measurements. These effects and the 
method by which we account for them are described in Section \ref{sec:nonideal}.

\section{Calibration}
\label{sec:calibration}
We next describe our map calibration procedures.  First, we discuss the relative calibration between detector TOD amplitudes.  Second, we describe how SPTpol maps are absolutely calibrated to those from SPT-SZ and \planck.  Finally, we describe our measurements of detector polarization angles and polarization efficiencies, which impact our $Q$ and $U$ map calibrations.

\subsection{Relative Calibration}
\label{sec:relcal}
Calibration of the detector response amplitude, or gain, across the array is particularly important 
for polarization-sensitive bolometric 
receivers. Because such instruments measure linear polarization by comparing the
intensity recorded on detectors with orthogonal polarization sensitivity, a difference in gain 
between orthogonal detectors will cause unpolarized radiation to appear polarized.  We monitor the relative gain among detectors---as well as changes in detector gain over 
time---through a combination of regular observations
of the galactic HII region RCW38 and regular observations of an internal chopped blackbody source.
The schedule of calibration observations and the analysis of the resulting data
are nearly identical to those used for SPT-SZ and
described in detail in \citet{schaffer11}; we summarize the salient features here.

A 45-minute observation of RCW38 (in which all detectors are scanned across the source multiple
times) is taken approximately every 20 hours, while one-minute observations of the internal source
are taken at least once per hour. Different detectors see the internal source with different illumination,
so the first step in the relative calibration pipeline is to assign a value for the effective temperature 
of the internal source for each detector. This value is based on the season average
of the ratio of that detector's response to the internal source and that detector's response to 
RCW38. Up to corrections for temperature drift of the source and atmospheric opacity, each detector's 
gain in a given CMB field observation is then
set by the combination of this effective temperature and the detector's response to the internal source
observation nearest the CMB field observation. We correct for any drifts in the internal source temperature 
using the ratio of response to an individual observation of the internal source to the season average of
that response, averaged over a detector module. Similarly, we correct for
changes in atmospheric opacity using the ratio of response to an individual observation of RCW38 
to the season average, again averaged over detector module.

\subsection{Absolute Calibration}
\label{sec:abscal}
For our absolute calibration, we compare temperature maps of the 100 \degs\ field made with SPTpol
data to maps of the same field made with SPT-SZ data. 
The SPT-SZ data has itself been calibrated by comparing the temperature power spectrum of the full 2500 \degs\ SPT-SZ 
survey to the published \planck\ power spectrum \citep{story13, planck13-16}. 
The uncertainty on this SPT-SZ-to-\planck\ calibration is estimated to be $1.2\%$ in temperature. 

We compare the two sets of maps by creating cross-power spectra. 
Specifically, we construct many sets of two half-depth maps of SPT-SZ data by splitting all single observations of the 
field into many sets of two halves and coadding. 
We also create one set of two half-depth SPTpol maps and a full-depth SPTpol map. 
We calculate the cross-spectrum between  the two half-depth SPTpol maps using a procedure similar to that described in Section \ref{sec:crossspectra}. 
For each semi-independent half-depth SPT-SZ map, we calculate the cross-spectrum between the half-depth SPT-SZ map and the full-depth SPTpol map, and we calculate the cross-spectrum ratio
\begin{equation}
r_{b,i} = \frac{\widehat{D}^\mathrm{\tiny{SPTSZ \ SPTpol}}_{b,i}}{\widehat{D}^\mathrm{\tiny{SPTpol \ SPTpol}}_b} \
\frac{B^\mathrm{SPTpol}_b}{B^\mathrm{SPTSZ}_b},
\end{equation}
where $\widehat{D}^{AB}_b$ is the binned cross-spectrum between maps $A$ and $B$ (details in Section \ref{sec:crossspectra}), 
$B^\mathrm{SPTpol}_b$ is the SPTpol beam averaged over an $\ell$-space bin
(and similar for SPT-SZ), the index $i$ runs over the different half-depth SPT-SZ maps used,
and the bins $b$ span the range $200 \le \ell \le 2200$ and have width $\Delta \ell = 50$.
All maps from both experiments used in this procedure have been created with identical observation strategy and filtering, so the filter transfer function effectively divides out of this ratio. 
For each bin, we calculate the mean and variance of $r_{b,i}$ across all SPT-SZ half-depth maps.
We check that the distribution of $r_b$ across all bins is consistent with a single underlying value, 
and we calculate the final ratio as the inverse variance-weighted mean ratio over all bins. 
In the power spectrum pipeline described in Section \ref{sec:ps}, we multiply all SPTpol maps by this final ratio before calculating cross-spectra.

The statistical uncertainty on the final SPT-SZ/SPTpol calibration ratio is given by the inverse of the square root of the total weight in all bins and is calculated to be $0.5 \%$. 
In the procedure described above, we have treated the SPTpol maps as noiseless,
because we expect the uncertainty in the ratio to be dominated by the noise in the SPT-SZ maps, 
which is roughly three times the noise level of the SPTpol data on this field.\footnote{We only use data from the 2008 SPT-SZ observations of this field. 
Data were taken on this field in 2010 as well, but with an observation strategy that makes it more difficult to match the filter transfer function to SPTpol.}
We have repeated the calculation using many sets of SPTpol half maps, and the change in the final uncertainty is minimal. 
The $1.3\%$ temperature calibration uncertainty used in the cosmological fits described in Section \ref{sec:cosmomc} is the quadrature sum of the uncertainties of the SPTpol-to-SPT-SZ and SPT-SZ-to-\planck\ calibrations.

\subsection{Polarization Calibration}
\label{sec:polcal}
Accurate reconstructions of $T$, $Q$, and $U$ maps require precisely measured polarization angles and polarization efficiencies for each detector.
SPTpol detectors operating at 150~GHz are arranged on the focal plane with nominal orientations 
from 0$^\circ$ to 180$^\circ$ in steps of 15$^\circ$.  
To measure the true detector polarization angles (including potential effects from telescope optics)
and to measure polarization efficiency, we perform a series of dedicated observations of a polarized calibration source located three km away from the telescope.  
The polarization calibrator consists of a chopped thermal source located behind two wire grid polarizers.  The grid closest to the thermal source is stationary and is used to establish a known polarization, while the second grid is rotated to modulate the polarization signal.
In order to avoid saturating detectors with the low-elevation atmosphere,
the source is placed in the middle of a 7\,m $\times$ 7\,m reflecting panel that redirects beams to the sky at an elevation of 60$^\circ$.  
A 2\,m-high wooden fence is installed half way between the source and the telescope to block reflections off the ground.

For each pair of detectors in a pixel, the telescope is pointed such that the source lies at the center of that pixel's beam.
The rotating polarizer is then stepped back and forth from  0$^\circ$ to 165$^\circ$ in 15$^\circ$ steps, 
and the detector response to the chopped signal as a function of the angle of the rotating grid is measured.  
We fit the response as a function of rotating grid angle to a model that has the detector polarization angle and polarization efficiency as free parameters.
This procedure is repeated for all detectors on the focal plane, with multiple measurements per detector where possible. 
We use these observations to establish distributions of measured angles for each grouping of detectors in a given module with a particular nominal angle.  
For detectors without a direct angle measurement that pass data quality cuts ($\sim 40\%$ of the array), we assign the median value from the appropriate distribution.  The same process is employed for deriving the polarization efficiencies.

Only observations with a statistical uncertainty of $< 2^\circ$ on the alignment angle and $< 5\%$ on the polarization efficiency are used for the polarization calibration analysis.  Additional cuts are made on the goodness-of-fit of the best-fit model and detector linearity, both of which cut a small fraction of observations relative to the parameter uncertainty cuts.  For the observations remaining after these cuts, the median statistical error on the fits to the polarization angle and efficiency are 0.46$^\circ$ and 0.8\%, respectively, per detector.  The mean difference between the measured and nominal alignment angles for all observations passing data quality cuts is $-1.0^\circ$.  The mean polarization efficiency is $97\%$.

To estimate the systematic uncertainty on our alignment angles, we repeat the measurements for a subset of approximately 10 well-behaved detectors in several different experimental configurations.
First, to test for any sensitivity to the beam shape and our ability to focus on the external calibration source, we repeat the measurements with the telescope intentionally de-focused.  Second, to test for sensitivity to reflections of the calibration signal off the ground, we repeat the measurements after removing the wooden fence located half way to the source.
Third, to test for our sensitivity to in-pixel crosstalk, we perform observations with only a single bolometer in each pixel biased in the superconducting transition.  Finally, to check that our measurements are robust to a different function of source output power versus rotating polarizer angle, we remove the fixed polarizing grid inside the calibration source.  For this last test, observations are repeated for all detectors rather than the subset of well-behaved detectors.  We compare the measured angles from each of the above tests to those from observations in the standard configuration and find that resulting mean differences are consistent with zero, using an error on the mean calculated with a conservative per-detector, per-observation angle uncertainty of 1.5$^\circ$.
Using this same conservative value for the per-observation uncertainty, we estimate the total systematic error for each detector's alignment angle to be 1$^\circ$.  This systematic uncertainty is negligibly small for the $TE$ and $EE$ power spectrum measurements presented here, as it would result in a $\lesssim 0.1\%$ change in the amplitude of these spectra.  The impact of the yet smaller statistical uncertainty was assessed using simulations, as described in Section~\ref{sec:nonideal}, and was also found to be negligibly small.  Any error in the measured polarization efficiencies will leak temperature anisotropy into polarization anisotropy, and we address this effect in Section \ref{sec:tp}.

\section{Bandpower Covariance Matrix}
\label{sec:cov}
The bandpower covariance matrix quantifies the uncertainties in individual bandpowers and the correlations between bandpowers.  We include covariance between $EE$ and $TE$ bandpowers, 
giving the covariance matrix a $2\times2$ block structure.  The ``on-diagonal" blocks are auto-covariances ($TE \times TE$ and $EE \times EE$), while the two ``off-diagonal" blocks contain the cross-covariance ($TE \times EE$).  
The covariance matrix includes contributions from noise variance, sample variance due to the finite number of modes measured in any given $\ell$-space bin, uncertainties in the instrument beam, and calibration uncertainty.

Sample variance for the auto-covariance blocks is estimated directly from the variance in the set of 782 simulated realizations of the SPTpol observations.
As described in \citet{lueker10}, the noise term is calculated from the variance in the measured cross-spectra $\widehat{D}^{AB}_b$.  The initial estimates of the sample and noise covariances $\widehat{\textbf{C}}_s$ and $\widehat{\textbf{C}}_n$ are biased, and we de-bias them using
\begin{equation}
\textbf{C}^{x,AB \times AB}_{bb^\prime} = \left(K^{-1}\right)_{bd}  \widehat{\textbf{C}}^{x,AB \times AB}_{dd^\prime} \left(K^{-1}\right)_{d^\prime b^\prime},
\end{equation}
where $AB \in \{TE,\,EE\}$, $x \in \{\mathrm{s},\,\mathrm{n}\}$ for sample and noise covariance, respectively, and recall the $K$ matrix was defined in Section \ref{sec:UnbiasedSpectra}.
  
For reasons detailed in \citet{lueker10}, the signal-to-noise on off-diagonal elements is low.  Thus, we condition the off-diagonal elements of each block in the covariance matrix.  We first calculate the correlation matrix $\pmb{\rho}_{ii^\prime}$ for each auto-covariance block, and set all elements $\ell > 400$ from the diagonal to zero.  The shape of the correlation matrix is determined by mode-mode coupling, and is therefore only a function of distance from the diagonal.  Thus, all remaining off-diagonal elements are replaced with the average of those elements at a fixed distance from the diagonal,
\begin{equation}
\label{eqn:covcond}
\pmb{\rho}^\prime_{ii^\prime}=\frac{
\sum_{i_1-i_2=i-i^\prime}\pmb{\rho}_{i_1i_2}}{\sum_{i_1-i_2=i-i^\prime} 1}.
\end{equation}
The conditioned correlation matrix is then transformed back into the corresponding auto-covariance block.

The $TE \times EE$ cross-covariance block is challenging to calculate.  
Since off-diagonal correlations between spectra are inherently small, and we generate a limited number of map bundles with which to estimate noise variance, calculating this block as described above yields low signal-to-noise matrix elements.
Instead, we construct the $TE \times EE$ block from the pre-conditioned and de-biased auto-covariance matrices.  Assuming covariances between different spectra are related by the theoretical expectations given by \citet{zaldarriaga97b}, we define the $TE \times EE$ diagonal elements as an algebraic combination of the auto-covariance diagonals calculated above.  The signs of the diagonal elements match those of the measured $TE$ bandpowers.  The off-diagonal shape of the matrix is defined by its correlation matrix, which we set as the mean of the $TE \times TE$ and $EE \times EE$ correlation matrices.  To determine the signs of off-diagonal matrix elements, we studied the covariances between $TE$ and $EE$ spectra generated from simulated maps with a simple cosine apodization mask applied.  We found that signs in the simulated $TE \times EE$ cross-covariance propagated perpendicularly away from the diagonal.  Off-diagonal elements between two diagonal elements received the sign of the averaged diagonal elements.  Changing apodization masks in the simulations did not alter this behavior.  Therefore, we apply the same off-diagonal sign propagation to the constructed $TE \times EE$ matrix.

Additional bin-to-bin covariance is generated due to uncertainties in the measurement of the beam function $B_{\ell}$.  
A ``beam correlation matrix" is first constructed,
\begin{equation}
\pmb{\rho}^{beam}_{ij} = \left(\frac{\delta D_i}{D_i}\right) \left(\frac{\delta D_j}{D_j}\right)
\end{equation}
where
\begin{equation}
\frac{\delta D_i}{D_i} = 1-\left(1+\frac{\delta B_i}{B_i}\right)^{-2},
\end{equation}
and $\frac{\delta B_i}{B_i}$ comes from the uncertainty in our measurements of Mars.  The beam correlation matrix is then converted to a covariance matrix via
\begin{equation}
\textbf{C}^{beam}_{ij} = \pmb{\rho}^{beam}_{ij}D_{i}D_{j}.
\end{equation}
Here $D_i$ and $D_j$ are drawn from the set of $TE$ or $EE$ bandpowers.

Finally, we may add covariance from uncertainty in our map calibration.  Both temperature and polarization maps get multiplied by a temperature calibration factor, $T_\mathrm{cal}$,  while polarization maps are also multiplied by a polarization calibration factor, $P_\mathrm{cal}$.  In practice, we choose to keep these parameters free during cosmological fitting, as discussed in Section \ref{sec:cosmomc} below, so we do not include covariance from calibration uncertainty by default.  However, we can generate a total calibration uncertainty $\pmb{\epsilon}_{XY}$ from the uncertainties on $T_\mathrm{cal}$ and $P_\mathrm{cal}$ for each block in the covariance matrix, where $X$ and $Y$ are either $TE$ or $EE$. 
The calibration covariance is then defined as
\begin{equation}
\textbf{C}^{cal}_{ij} = \pmb{\epsilon}_{XY}D_{i}D_{j},
\end{equation}
where again $D_i$ and $D_j$ are $TE$ or $EE$ bandpowers corresponding to $X$ and $Y$.

\section{Tests for Systematic Errors}
\label{sec:systematicTests}

\subsection{Null Tests}
\label{sec:nulltests}
We perform a suite of null, or jackknife, tests to check the internal consistency of our measurement.
A jackknife test entails dividing the data into two sets using a metric that might be associated with a systematic effect.
We pair each map bundle in the first set with a map bundle in the second set then difference the pairs of bundles to remove the common CMB signal.  We perform the cross-spectrum analysis on the resulting sets of null maps with the same procedure we use to determine the power spectra of the data.  Due to small changes in weights and filtering from observation to observation, the ``expectation spectra" for each null test are in general non-zero.  We calculate these expectation spectra by applying null tests to simulated maps.  For each null test, the measured power spectra should be consistent with the expectation spectra if the particular systematic effect being probed is not present above our noise level.  We test this by calculating the $\chi^2$ of the residual power relative to the expectation spectra in nine bins with $\Delta \ell=500$, from $\ell$ of 500 to 5000.  We calculate the probability to exceed (PTE) of this value of $\chi^2$ with nine degrees of freedom.

We present four map-based jackknife tests.
\begin{enumerate}
\item[-]  Left/Right: Difference maps are made by subtracting data in left-going scans from data in right-going scans.
Power that is different in left and right-going scans could be induced by asymmetric telescope scanning and elevation steps.  
\item[-] 1st half/2nd half: Difference maps are made by subtracting data from the first half of the observing season from data from the second half of the observing season.
This tests for systematic effects with a temporal dependence. 
Temporal variations in power could be caused by a calibration drift, time dependence of systematic signals, or the sun being above the horizon at the end of the season.
\item[-] Ground: Difference maps are made by dividing the maps based on potential ground contamination 
(using the same azimuthal range metric used in SPT-SZ power spectrum analyses, e.g., \citealt{shirokoff11})
and subtracting the worst half of maps from the best half.
Power from ground contamination could be caused by features on the horizon such as buildings near the telescope.
\item[-] Moon: 
Difference maps are made by subtracting data taken when the moon was above the horizon from data
taken when the moon was below the horizon.
Power from the moon might be picked up via sidelobes when the moon is above the horizon.
\end{enumerate}

The results of the tests are summarized in Table \ref{tab:jk1}.
The resulting PTE values for both the $TE$ and $EE$ spectra are roughly uniformly distributed between zero and one.   We conclude that there is no evidence for systematic bias from this suite of null tests.

\begin{table}
\begin{center}
\caption[]{Jackknife Tests}
\begin{tabular}{l|c|c}
\hline\hline
\rule[-2mm]{0mm}{6mm}
Test & $TE$ & $EE$ \\ 
\hline 
Left/Right & 0.58 & 0.06 \\
1st half/2nd half & 0.43 & 0.64 \\
Ground & 0.74 & 0.44 \\
Moon & 0.12 & 0.58 \\
\hline 
\end{tabular}
\tablecomments{
The results of the jackknife tests are quoted as the probability to exceed (PTE) the $\chi^2$ per degree of freedom for each test. }
\label{tab:jk1}
\end{center}
\end{table}

\subsection{Other Potential Systematic Effects}
\label{sec:othertests}

\subsubsection{Temperature to Polarization Leakage}
\label{sec:tp}
A variety of systematic effects can cause filtered versions of the sky temperature, $T$, to contaminate our estimates of the Stokes $Q$ and $U$ polarization.  
In the simplest case, the contamination is a scaled version of the temperature map.  
For example, the contaminated portion of the $Q$ map would be $Q^{\rm{contam}} = \epsilon^{Q}T$.  
As the fractional polarization of the CMB is small, even a small amount of leakage from $T$ can contaminate the $Q$ and $U$ signals; thus it is important to ensure that we correct for any such leakage.

We estimate the leakage parameters, $\epsilon^{P}$, where $P=\{Q,U\}$, using the cross-correlation between the temperature and polarization maps:
\begin{equation}
\hat{\epsilon}^{P} = \frac{\sum_{\ell, \phi_\ell} C^{TP}_{\ell, \phi_\ell} }{\sum_{\ell, \phi_\ell} C^{TT}_{\ell, \phi_\ell} }.
\end{equation}
Here, the subscripts $(\ell, \phi_\ell)$ denote the radius and azimuthal angle, respectively, of the spectra in 2-D Fourier space.

We measure the cross-spectra, $C^{TP}_\ell$ and $C^{TT}_\ell$, using two maps, 
each of which contains half of the full set of data analyzed here, 
and we evaluate the sums across the multipole range $500<\ell<2500$.
We find $\hat{\epsilon}^{Q}=0.0105$ and $\hat{\epsilon}^{U}=-0.0152$.  To correct for this leakage we subtract the appropriately scaled temperature map from each polarization map.  We find that the $TE$ and $EE$ bandpowers shift by an amount that is small compared to their
uncertainties when these corrections are applied, and conclude that any
additional uncertainty caused by the $\pm0.0015$ uncertainty in the
leakage parameters can be ignored.
We also find that performing the same procedure on maps made using leakage-free, 
simulated observations introduces a negligible bias in the reconstructed $C^{TE}_\ell$ and $C^{EE}_\ell$ spectra.

The temperature-to-polarization contamination described above is the so-called ``monopole'' leakage, 
in which the contamination is simply a scaled version of the temperature map.  
However, more complicated forms of leakage could exist.  
For example, uncorrected pointing offsets introduce dipole temperature leakage, 
while differential beam ellipticity introduces quadrupole temperature leakage.  These higher-order leakage terms form a $TE$ ``leakage beam," $G_\ell^{TE}$.  
We estimate the leakage beam by calculating the cross-correlation between $T$ and $E$-mode polarization maps of Venus observations:
\begin{equation}
G_\ell^{TE} = \frac{\sum_{\phi_\ell} \left( C^{TE}_{\ell, \phi_\ell} \right)_{\mathrm{Venus}} }{\sum_{\phi_\ell} \left( C^{TT}_{\ell, \phi_\ell} \right)_{\mathrm{Venus}} }.
\end{equation}
Observations of Venus, which is effectively a point source given the measured beam FWHM, have high signal-to-noise out to a radial distance of $\sim 10$ 
arcminutes, so this procedure captures the features of the leakage beam out to this angular scale.
To remove leaked systematic power in our $TE$ spectrum we subtract a copy of our measured $TT$ spectrum scaled by the leakage beam,
\begin{equation}
C_{\ell,\mathrm{corrected}} ^{TE} = C_{\ell,\mathrm{uncorrected}} ^{TE} - G_\ell^{TE} C_\ell^{TT}.
\end{equation}
Before applying this correction we find significant ($\sim 4 \sigma$) evidence for roughly constant-in-$C_\ell$ power 
beyond that expected from \LCDM in our $TE$ spectrum (see Section \ref{sec:cosmomc} for details).  After applying the $TE$ leakage beam correction the significance drops to $1.4 \sigma$.  
We conclude that, after the $TE$ leakage beam correction, our $TE$ spectrum is free of statistically significant leakage power.  
Since the leakage power enters into the $EE$ spectrum in a quadratic sense (i.e., with a prefactor of $[G_\ell^{TE}]^2$), we also conclude that the higher-order (non-monopole) temperature-to-polarization leakage in our $EE$ spectrum is negligible in this analysis.

\subsubsection{Detector Non-Idealities}
\label{sec:nonideal}

Our map-making procedure assumes that our detectors behave ideally
and have been perfectly characterized.
We know of several ways in which the true behavior of the detectors, or our knowledge of 
that behavior,
violates this assumption.
We estimate the effects on the $EE$ and $TE$ power spectra from each of these non-idealities
individually through simulations. If necessary, we correct our power spectrum estimates using
the results of these simulations.

The response of the detectors to sky signals depends slightly on the amount of optical power on the detector.  The column depth of atmosphere seen by the detectors changes with observing elevation,
resulting in an elevation-dependent detector responsivity.  
By taking measurements with an internal calibration source at many elevations, we determine 
the change in responsivity as a function of elevation for each detector individually.

There is also some low-level electrical crosstalk between detectors.
The observations of RCW38 discussed in Section \ref{sec:relcal} are used to characterize the crosstalk.
Crosstalk manifests itself in these observations as duplicate copies of RCW38 in 
a single detector's map of the source. A model profile is constructed for each detector with one copy of 
RCW38 at the center of the map and duplicates at the relative locations of all other detectors,
with the amplitudes of the duplicates as free parameters.
This model profile is fit to each detector's map to determine the crosstalk matrix $X_{ab}$, which encodes how
signal from detector $b$ leaks into the TOD of detector $a$.

The other detector non-ideality we investigate through simulations is imperfect knowledge of the detector polarization
angles. The uncertanties in the angle measurements are estimated as described in Section \ref{sec:polcal}.

Once these non-idealities are characterized, their effect on the output power spectrum is investigated using the 
simulation pipeline described in Section \ref{sec:transfer}.  The power spectra with these non-idealities included in the simulations are compared against the power spectra from simulations assuming ideal detector operation.   
The power spectrum errors introduced by elevation-dependent responsivity and detector angle uncertainty are below one tenth of a sigma for each $\ell$-space bin and are randomly scattered. 
Their total effect on our cosmological fits are negligible, so we ignore these two non-idealities.

The effect of electrical crosstalk, while still smaller than our bandpower uncertainties, is strongly correlated between $\ell$-space bins, and we correct our final bandpower estimates for this effect.  The bulk of 
the effect is due to the fact that the beam estimate we use to relate our biased power spectrum estimates
to the true power spectrum (Equation \ref{eqn:kdef}) is measured in temperature-only maps. The 
electrical crosstalk in SPTpol is predominantly negative, and the effect of crosstalk on the composite 
beam measured in temperature-only maps made from the data of many detectors
is to impart negative lobes at a distance
away from the main lobe equal to the mean crosstalk partner separation. (For SPTpol, this distance is 
roughly two arcminutes.) This will be the correct effective beam for any temperature-only maps made 
with the same detectors and weighting. For polarization maps, however, the mean effect of crosstalk
is zero, unless the amplitude of the crosstalk is correlated with detector polarization angle, which we 
see no evidence of in SPTpol data. This means
that when we use the beam measured from temperature maps with cross-talk in the 
polarized power spectrum estimation, we are incurring a multiplicative
bias related to the ratio of this beam to the true, non-cross-talk-biased beam. 

We estimate the exact form of this bias in simulations and find that the bias imparts a roughly linear tilt on both spectra.  The tilt ranges from $+2\%$ to $-4\%$ for $\ell=500 - 5000$ in the $EE$ spectrum and from $+1\%$ to $-2\%$ in the $TE$ spectrum across the same multipole range.  After we correct for this bias, \LCDM parameter values shift by less than $0.1\,\sigma$ and the limit on residual Poisson power in our $EE$ spectrum moves by $0.4\,\sigma$ (see Section \ref{sec:ptsrc} for details).  The uncertainty on our simulation-based estimate of this multiplicative bias is roughly 30\%.  The effect of this bias uncertainty on parameter estimates is far below our statistical uncertainties on all parameters, so we ignore it in our final cosmological fits.

\subsubsection{Sensitivity of the Analysis to Cosmological Model}
We test the sensitivity of our analysis to differences between the model we assume for
sky power and the actual sky power we measure.  To accomplish this, we create simulated maps with an input spectrum shifted by $\Delta \ell=10$ from the \LCDM model spectrum assumed in the calculation of the transfer function in Section \ref{sec:transfer}.  This approximates shifting the angular scale of the sound horizon at matter-radiation decoupling $\theta_\mathrm{s}$.  With the detection of several acoustic peaks in both the $TE$ and $EE$ spectra, $\theta_\mathrm{s}$ is one of the parameters to which the SPTpol dataset is most sensitive.  Using our standard pipeline and transfer function, we then calculate the power spectra of these simulated maps and compare them with the input spectra.  
From this test we recover the $\ell$-shifted input spectra to well within the stated statistical uncertainty.

\begin{figure*}
\begin{center}
\includegraphics[width=0.5\textwidth]{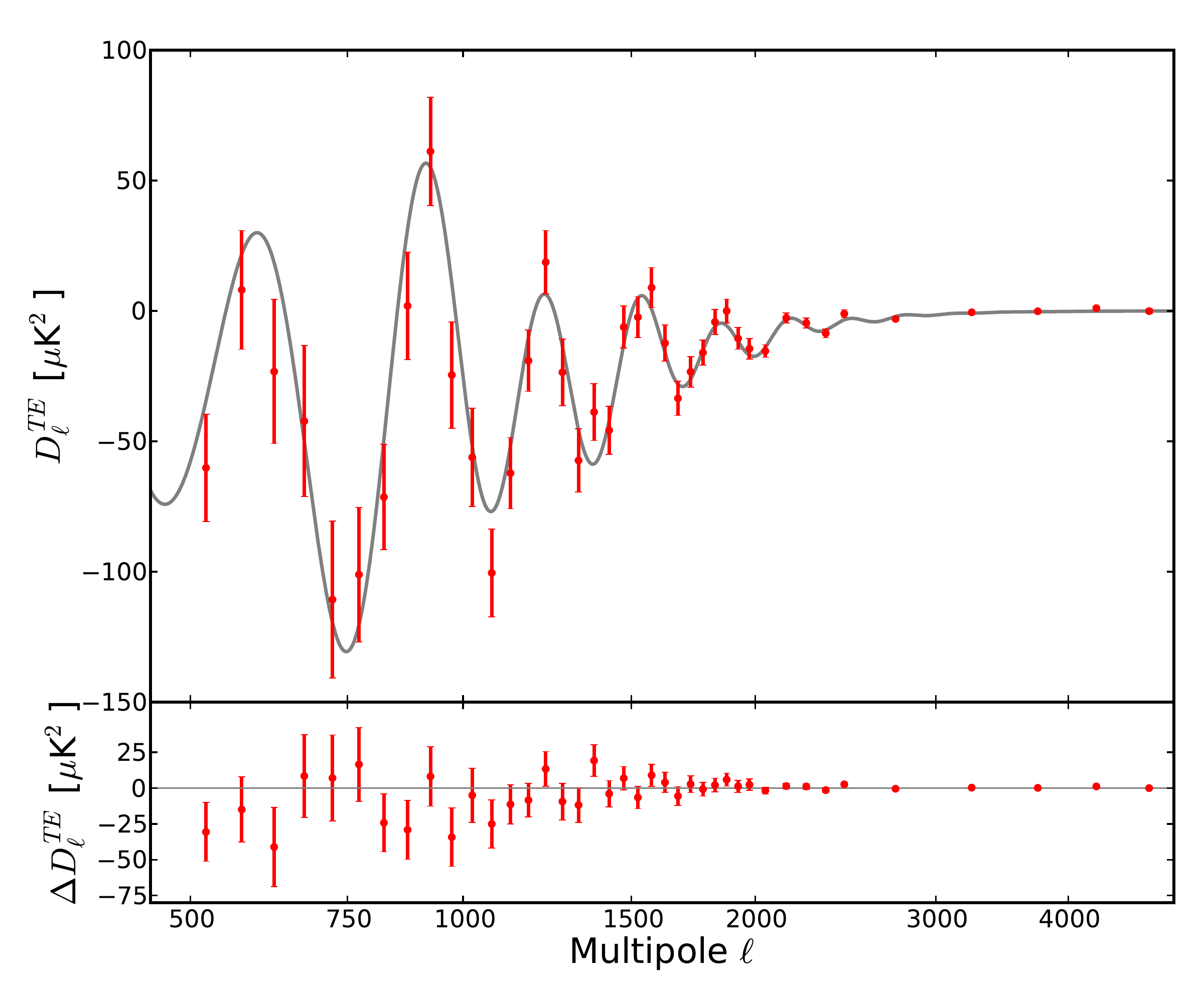}\includegraphics[width=0.5\textwidth]{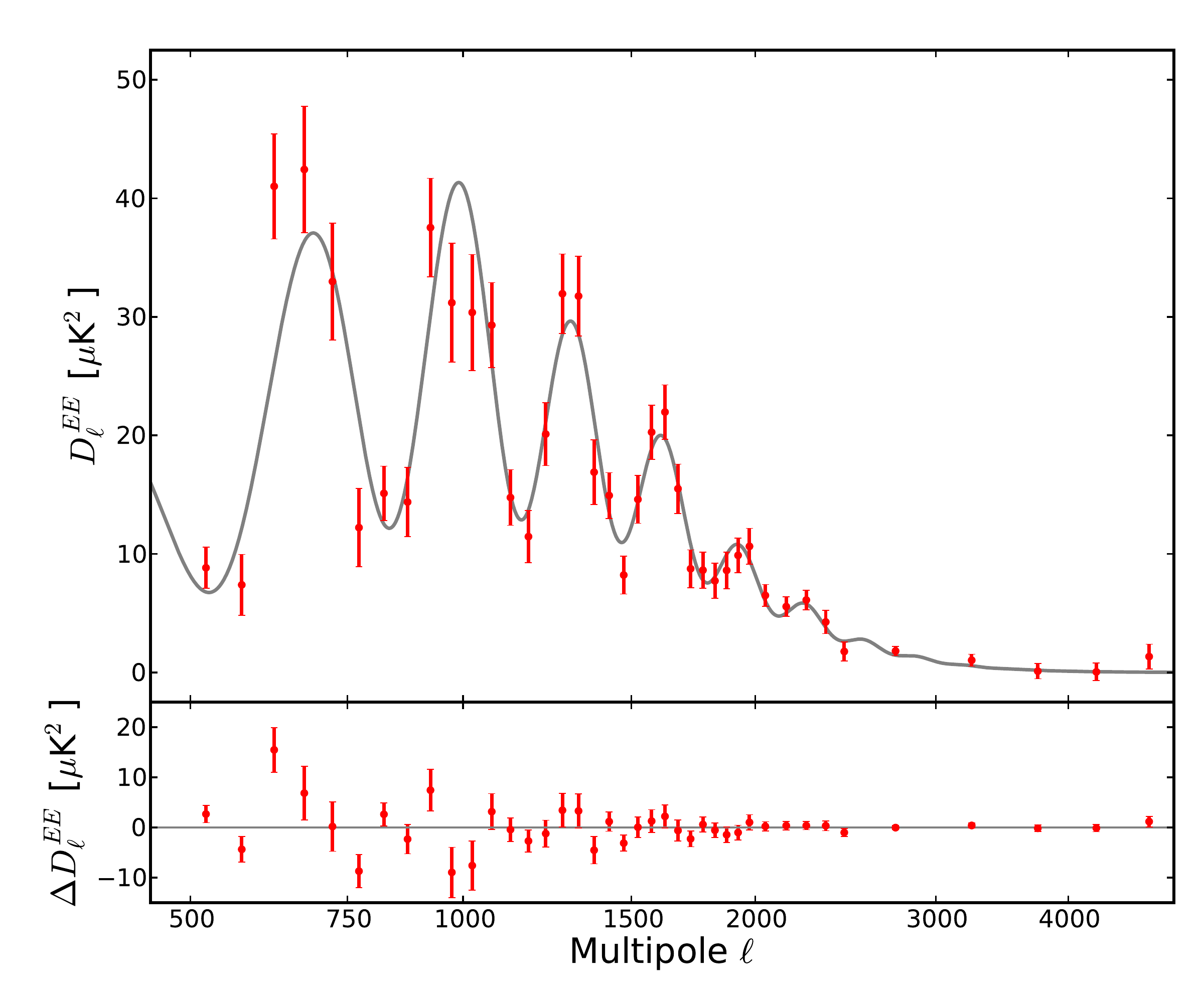}

\end{center}
\caption{The SPTpol $TE$ (Left) and $EE$ (Right) power spectra.  
The solid gray lines are the \textsc{Planck+SPT-SZ+SPTpol} best-fit \LCDM{} model described in Section \ref{subsec:sptpol}.  The x-axis is scaled to $l^{0.5}$.  Residuals, $\Delta D_{\ell}$, to the best-fit model are plotted in the sub-panels.  Bandpower error bars include sample and noise variance. 
}
\label{fig:spectra_sptpol}
\vspace{0.1in}
\end{figure*}

\begin{figure*}
\begin{center}
\includegraphics[width=0.98\textwidth]{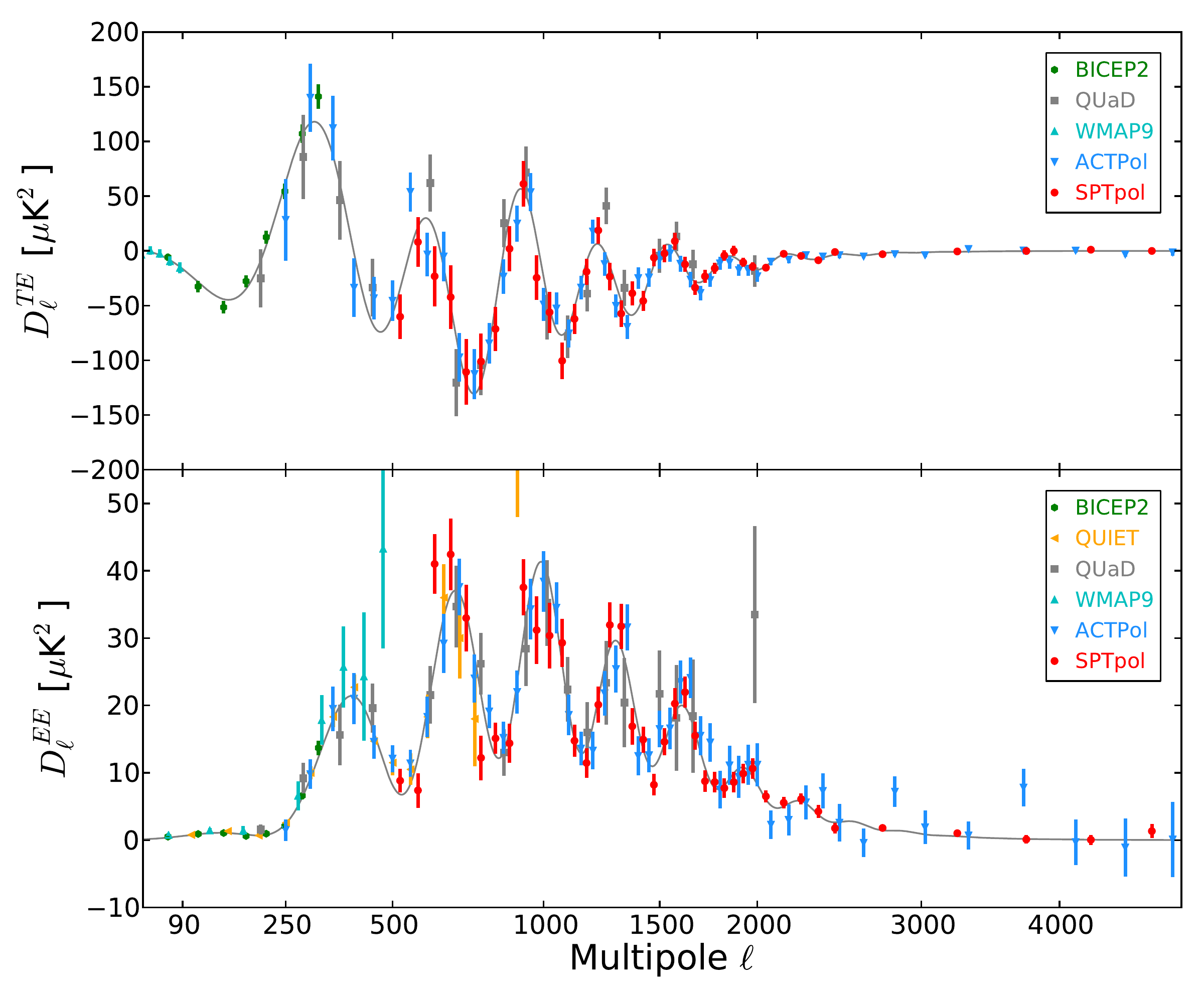}
\end{center}
\caption{$TE$ (top) and $EE$ (bottom) power spectrum measurements.
In addition to SPTpol, we plot data from BICEP2 \citep{bicep2a}, WMAP9 \citep{hinshaw13}, QUIET W-band \citep{quiet12}, QUaD \citep{Brown09}, and ACTpol \citep{naess14}.  
The solid gray lines are the \textsc{Planck+SPT-SZ+SPTpol} best-fit \LCDM{} model described in Section \ref{subsec:sptpol}.  The x-axis is scaled to $l^{0.5}$.
}
\label{fig:spectra_all}
\vspace{0.1in}
\end{figure*}

\section{Bandpowers}
\label{sec:bandpowers}

We present bandpowers and uncertainties for the $TE$ and $EE$ spectra at 150 GHz in Figure \ref{fig:spectra_sptpol} and Table \ref{tab:TEEE}.  The SPTpol bandpowers span the range $500 < \ell\leq 5000$.  In Figure \ref{fig:spectra_all}, we plot the SPTpol bandpowers with recent measurements by other experiments.  In both plots the solid gray lines are the \textsc{Planck+SPT-SZ+SPTpol} best-fit \LCDM{} model described in Section \ref{subsec:sptpol} below.  The plotted SPTpol errors are the square root of the diagonal elements of the relevant auto-covariance block and do not include beam or calibration uncertainties.  
As described in the next section, temperature and polarization absolute calibration are free parameters in the cosmological fits, and the bandpowers are corrected by the best-fit values for these parameters 
and for the mean effect of crosstalk (Section \ref{sec:nonideal}).
The $TE$ bandpowers have also been corrected for temperature-to-polarization leakage as described in Section \ref{sec:tp}.  The bandpowers are available at the SPT website\footnote{http://pole.uchicago.edu/public/data/crites14/} along with the covariance matrix and bandpower window functions.

The $\ell$-space bins used to calculate SPTpol bandpowers have three widths: 
$\delta\ell$ = 50 from 501 $\leq$ $\ell$ $\leq$ 2000, $\delta\ell$ = 100 from 2001 $\leq$ $\ell$ $\leq$ 2500, 
and $\delta\ell$ = 500 from 2501 $\leq$ $\ell$ $\leq$ 5000.  
The third through eighth peaks of the $EE$ power spectrum are measured with high signal-to-noise.  
To date, this is the highest-fidelity measurement of the photon-diffusion-damped region of the $EE$ and $TE$ power spectra.

\begin{table}[ht!]
\begin{center}
\caption[SPTpol $TE$ and $EE$ Bandpowers and Bandpower Errors]{SPTpol $TE$ and $EE$ Bandpowers and Bandpower Errors}
\small
\begin{tabular}{cr | cc | cc}
\hline\hline
\rule[-2mm]{0mm}{6mm}
$\ell$ range&$\ell_{\rm eff}$ &$D_{\ell}^{TE}$ & $\sigma^{TE}$ & $D_{\ell}^{EE}$ & $\sigma^{EE}$ \\
\hline

501  -  550  &  521  &  -60.2  &  20.5  & 8.8  &  1.7  \\
551  -  600  &  572  &  8.1  &  22.8  & 7.4  &  2.6  \\ 
601  -  650  &  622  &  -23.3  &  27.6  & 41.0  &  4.4  \\ 
651  -  700  &  672  &  -42.3  &  28.9  & 42.4  &  5.3  \\ 
701  -  750  &  722  &  -110.7  &  30.1  & 33.0  &  4.9  \\ 
751  -  800  &  772  &  -101.1  &  25.8  & 12.2  &  3.3  \\ 
801  -  850  &  822  &  -71.4  &  20.2  & 15.1  &  2.3  \\ 
851  -  900  &  872  &  1.9  &  20.6  & 14.4  &  2.9  \\ 
901  -  950  &  923  &  61.2  &  20.8  & 37.5  &  4.2  \\ 
951  -  1000  &  973  &  -24.6  &  20.4  & 31.2  &  5.0  \\ 
1001  -  1050  &  1023  &  -56.2  &  18.9  & 30.4  &  4.9  \\ 
1051  -  1100  &  1073  &  -100.5  &  16.8  & 29.3  &  3.6  \\ 
1101  -  1150  &  1123  &  -62.2  &  13.7  & 14.8  &  2.3  \\ 
1151  -  1200  &  1173  &  -19.1  &  11.8  & 11.5  &  2.2  \\ 
1201  -  1250  &  1223  &  18.7  &  12.2  & 20.1  &  2.7  \\ 
1251  -  1300  &  1273  &  -23.5  &  12.8  & 31.9  &  3.3  \\ 
1301  -  1350  &  1323  &  -57.3  &  12.1  & 31.8  &  3.4  \\ 
1351  -  1400  &  1373  &  -38.8  &  10.9  & 16.9  &  2.7  \\ 
1401  -  1450  &  1423  &  -45.7  &  9.2  & 14.9  &  1.9  \\ 
1451  -  1500  &  1473  &  -6.2  &  8.1  &  8.2  &  1.6  \\ 
1501  -  1550  &  1523  &  -2.4  &  7.9  & 14.6  &  2.0  \\ 
1551  -  1600  &  1573  &  8.9  &  7.7  & 20.3  &  2.3  \\
1601  -  1650  &  1623  &  -12.3  &  6.9  & 22.0  &  2.3  \\ 
1651  -  1700  &  1673  &  -33.5  &  6.6  & 15.5  &  2.1  \\ 
1701  -  1750  &  1723  &  -23.3  &  5.9  & 8.8  &  1.6  \\ 
1751  -  1800  &  1773  &  -16.0  &  4.9  & 8.6  &  1.5  \\ 
1801  -  1850  &  1823  &  -4.2  &  4.8  & 7.7  &  1.5  \\ 
1851  -  1900  &  1873  &  -0  &  4.5  & 8.6  &  1.5  \\ 
1901  -  1950  &  1923  &  -10.6  &  4.2  & 9.9  &  1.5  \\ 
1951  -  2000  &  1973  &  -14.5  &  4.0  & 10.6  &  1.5  \\ 
2001  -  2100  &  2047  &  -15.4  &  2.4  & 6.5  &  0.9  \\ 
2101  -  2200  &  2147  &  -2.7  &  2.0  & 5.6  &  0.8  \\ 
2201  -  2300  &  2247  &  -4.6  &  1.9  & 6.1  &  0.8  \\ 
2301  -  2400  &  2348  &  -8.6  &  1.7  & 4.3  &  1.0  \\
2401  -  2500  &  2448  &  -1.1  &  1.5  & 1.8  &  0.8  \\ 
2501  -  3000  &  2745  &  -3.07  &  0.55  & 1.81  &  0.40  \\ 
3001  -  3500  &  3246  &  -0.53  &  0.48  & 1.03  &  0.51  \\ 
3501  -  4000  &  3746  &  -0.13  &  0.55  & 0.11  &  0.64  \\ 
4001  -  4500  &  4246  &  1.05  &  0.69  & 0.05  &  0.74  \\ 
4501  -  5000  &  4747  &  -0.07  &  0.83  & 1.34  &  1.05  \\ 

\hline
\end{tabular}
\label{tab:TEEE}
\tablecomments{
The $\ell$-range, weighted multipole value $\ell_{\rm eff}$, bandpower $D_{\ell}$, and associated bandpower uncertainty, $\sigma$, of the SPTpol 150 GHz $TE$ and $EE$ power spectra.  Bandpowers and errors are given in units of $\microKsq$.
The errors are the square-root of the diagonal elements of the $TE$ and $EE$ auto-covariance matrices, and do not include beam or calibration uncertainties.
}
\end{center}
\end{table}

\section{Cosmological Constraints}
\label{sec:constraints}

\subsection{Estimating Cosmological Parameters}
\label{sec:cosmomc}
We obtain constraints on cosmological parameters with \textsc{CosmoMC}, a Markov Chain Monte Carlo (MCMC) package \citep{lewis02b}.  As in past SPT analyses \citep[e.g., ][]{hou14}, we have configured \textsc{CosmoMC} to use \textsc{PICO}\footnote{https://sites.google.com/a/ucdavis.edu/pico/} \citep{fendt07b,fendt07a} trained with \textsc{CAMB}.  To calculate the likelihood for the SPTpol bandpowers, we have written a new SPTpol-specific module for \textsc{CosmoMC}, which is also available on the SPT website.

The SPTpol likelihood introduces four nuisance parameters that are marginalized over when obtaining constraints on the six standard \LCDM{} parameters.  The first and second nuisance parameters are temperature and polarization calibration, $T_\mathrm{cal}$ and $P_\mathrm{cal}$, discussed above.  After correcting the calibration of the maps as discussed in Sections \ref{sec:abscal} and \ref{sec:polcal} we apply a Gaussian prior to $T_\mathrm{cal}$ centered on unity with a standard deviation of 0.013 and a uniform prior on $P_\mathrm{cal}$ between 0.95 and 1.15.  Expanding either limit of the prior does not alter our cosmological results.  In this calibration scheme, $P_\mathrm{cal}$ is degenerate with multiplicative biases that affect polarization but not temperature data.  The combination of $T_\mathrm{cal}$ and $P_\mathrm{cal}$ can simultaneously account for calibration uncertainty and any constant-in-multipole multiplicative bias in either temperature-plus-polarization or polarization data alone.  When fitting for and marginalizing over $T_\mathrm{cal}$ and $P_\mathrm{cal}$ we do not include calibration uncertainty in the bandpower covariance matrix.  However, a version of the covariance matrix that includes calibration uncertainty calculated using the 68\% limits for $T_\mathrm{cal}$ and $P_\mathrm{cal}$ in Table \ref{tab:all_lcdm} is available on the SPT website.

The third nuisance variable is a foreground term,  $D^{\mathrm{PS_{EE}}}_{3000}$, parameterizing the level of residual polarized power from unclustered (or ``Poisson'') point sources at $\ell = 3000$ in the $EE$ spectrum after masking
all sources above 50 mJy in unpolarized flux.  
The $\ell$ dependence of this signal is $D_{\ell} \propto \ell^2.$

The final SPTpol nuisance parameter is $\kappa$, the mean lensing convergence in the field.  As discussed in \citet{manzotti14}, a small patch of sky is lensed by modes larger than the patch itself such that the scale of anisotropies is dilated by lensing across the entire patch.  For surveys with relatively small sky coverage such as the SPTpol 100d field, ignoring this effect can potentially lead to a non-negligible bias on the angular scale of the sound horizon at recombination, $\theta_\mathrm{s}$.  To account for this effect, we alter the theoretical spectra entering our likelihood calculation, which are functions of parameters $\textbf{p}$, to have dependence on $\kappa$,  
\begin{equation}
\hat{C}_\ell^{XY} \left( \mathbf{p};\kappa \right) = C_\ell^{XY} \left( \mathbf{p}\right) - \frac{\partial \ell^2 C_\ell^{XY} \left( \mathbf{p}\right)}{\partial \ln \ell}\frac{\kappa}{\ell^2},
\end{equation}
as suggested by \citet{manzotti14}.  We apply a Gaussian prior to $\kappa$ centered on zero with a standard deviation of $\sigma_{\kappa} = 2.45\times 10^{-3}$, which is the rms fluctuation in $\kappa$ across a 100 deg$^2$ circular field for the flat \LCDM Planck cosmology considered in \citet{manzotti14}.

To quantify the level of residual temperature-to-polarization leakage after correcting for monopole
and higher-order leakage terms (see Section \ref{sec:tp}), 
we include an extra nuisance parameter, $D^{\mathrm{PS_{TE}}}_{3000}$, which is defined analogously to the $EE$ foreground parameter above.  
Note that the expectation value for $TE$ from point sources is zero, even for a single source
\citep[e.g.,][]{tucci04}, so this parameter is only used to quantify residual $T$ to $P$ leakage.
For all of the cosmological fits discussed below, $D^{\mathrm{PS_{TE}}}_{3000}$ is fixed at zero.  

When fitting a cosmological model to the SPTpol bandpowers we also include two external datasets.  
In particular, we consider measurements of the CMB $TT$ spectrum from \planck\ \citep{planck13-16} as well as the 2500 deg$^2$ \textsc{SPT-SZ} survey \citep{story13}.  We note that the \planck, \textsc{SPT-SZ}, and SPTpol likelihoods treat foregrounds independently.  Given that the 100d SPTpol field is only a small fraction of the area surveyed by \textsc{SPT-SZ}, and both the SPTpol and \textsc{SPT-SZ} regions are small compared to the full sky surveyed by \planck, 
we also ignore any correlations between experimental results due to shared sky.

\subsection{Consistency with \LCDM}
\label{subsec:sptpol}
We check that the SPTpol dataset is consistent with the \LCDM model.  To quantify the goodness of fit, we calculate the $\chi^2$ between the \textsc{Planck+SPT-SZ+SPTpol} best-fit \LCDM model and the SPTpol $TE$ and $EE$ bandpowers and errors scaled by the best-fit calibration parameters ($T_\mathrm{cal}=0.992$ ; $P_\mathrm{cal}=1.047$).  (SPTpol bandpowers re-scaled by these calibration parameters are plotted in Figure \ref{fig:spectra_sptpol} along with their residuals to the \textsc{Planck+SPT-SZ+SPTpol} best-fit model).  The resulting $\chi^2$ is 95.1 with 80 total bandpowers.  While there are ten free parameters in the fit (six for \LCDM + four SPTpol nuisance parameters), the \textsc{Planck} and \textsc{SPT-SZ} datasets effectively fix the \LCDM parameters; with an observing area of only 100~deg$^2$, the SPTpol bandpowers have large sample variance over the range of multipoles that best constrain the standard \LCDM parameters.  Consequently, there are effectively only four free parameters and therefore 76 degrees of freedom.  This translates to a PTE of 0.07.  If instead we fix the \LCDM parameters to the best-fit values mentioned above and only fit for the SPTpol nuisance parameters, there are exactly four free parameters.  In this case the $\chi^2$ and PTE are unchanged.  We conclude the SPTpol bandpowers are adequately fit by the standard \LCDM model and proceed to consider joint cosmological constraints.

\subsection{$\mathrm{\LCDM{}}$ Constraints}
\label{subsec:lcdm}
Table \ref{tab:all_lcdm} summarizes the results of parameter fits to the standard flat \LCDM model with and without the inclusion of SPTpol data.  As in \citet{planck13-16}, our parameterization of \LCDM uses the approximate angular size of the sound horizon $\theta_{\mathrm{MC}}$ as calculated by \textsc{CosmoMC} instead of the true angular size $\theta_{\mathrm{s}}$.  We quote the amplitude of the spectrum $\ln(10^{10}A_{\mathrm{s}})$ at a pivot scale of $k_0 = 0.05$~Mpc$^{-1}$.  Constraints improve slightly when combining SPTpol with \textsc{Planck+SPT-SZ} and median parameter values move no more than $0.3\,\sigma$.

\begin{table}                                                                                                                                                                                                  

\begin{center}
\caption{\LCDM constraints}

\small

\begin{tabular}{c|c|c}
\hline\hline
\multicolumn{1}{c|}{Parameter} & \multicolumn{2}{c}{Dataset} \\
& \textsc{Planck+SPT-SZ} & \textsc{Planck+SPT-SZ}  \\
&  & \textsc{+SPTpol}  \\
\hline

\multicolumn{2}{l}{ Free } \\
$\Omega_{\mathrm{b}}h^2$ & $0.02207 \pm 0.00027$ & $0.02203 \pm 0.00026$ \cr
$\Omega_{\mathrm{c}}h^2$ & $0.1189 \pm 0.0025$ & $0.1185 \pm 0.0024$ \cr
$100\theta_{\mathrm{s}}$ & $1.04168 \pm 0.00058$ & $1.04164 \pm 0.00056$ \cr
$n_{\mathrm{s}}$ & $0.9597 \pm 0.0068$ & $0.9593 \pm 0.0067$ \cr
$\ln(10^{10}A_{\mathrm{s}})$ & $3.077 \pm 0.024$ & $3.070 \pm 0.024$ \cr
$\tau$ & $0.084 \pm 0.013$ & $0.081 \pm 0.012$ \cr
\hline

\multicolumn{2}{l}{ Derived } \\
$\Omega_{\mathrm{\Lambda}}$ & $0.692 \pm 0.015$ & $0.693 \pm 0.015$ \cr
$\sigma_8$ & $0.820 \pm 0.012$ & $0.816 \pm 0.012$ \cr
$H_0$ & $67.8 \pm 1.1$ & $67.9 \pm 1.1$ \cr
\hline

\multicolumn{2}{l}{ Nuisance } \\
$T_{\mathrm{cal}}$ & ---  & $0.992 \pm 0.012$ \cr
$P_{\mathrm{cal}}$ & ---  & $1.048 \pm 0.017$ \cr
$D^{\mathrm{PS_{EE}}}_{3000}$ & ---  & $<\,0.40\,\mu$K$^2$ at 95$\%$ \cr
$100\kappa$ & ---  & $0.047 \pm 0.168$ \cr
\hline

\hline
\end{tabular}
\label{tab:all_lcdm}

\tablecomments{
Median fits and symmetric 68\% limits.  Here \textsc{Planck} refers to \planck\ $TT$ bandpowers \citep{planck13-16} plus \textsc{WMAP9} polarization \citep{hinshaw13}.  For $D^{\mathrm{PS_{EE}}}_{3000}$ we quote the 95\% confidence upper limit and note that all sources above 50\,mJy in unpolarized flux have been masked in the analysis.}

\end{center}                                                                                                                                                                                          
\end{table}

\subsection{Constraints on Polarized Power from Extragalactic Sources}
\label{sec:ptsrc}

\begin{figure}
\begin{center}
\includegraphics[width=0.4\textwidth]{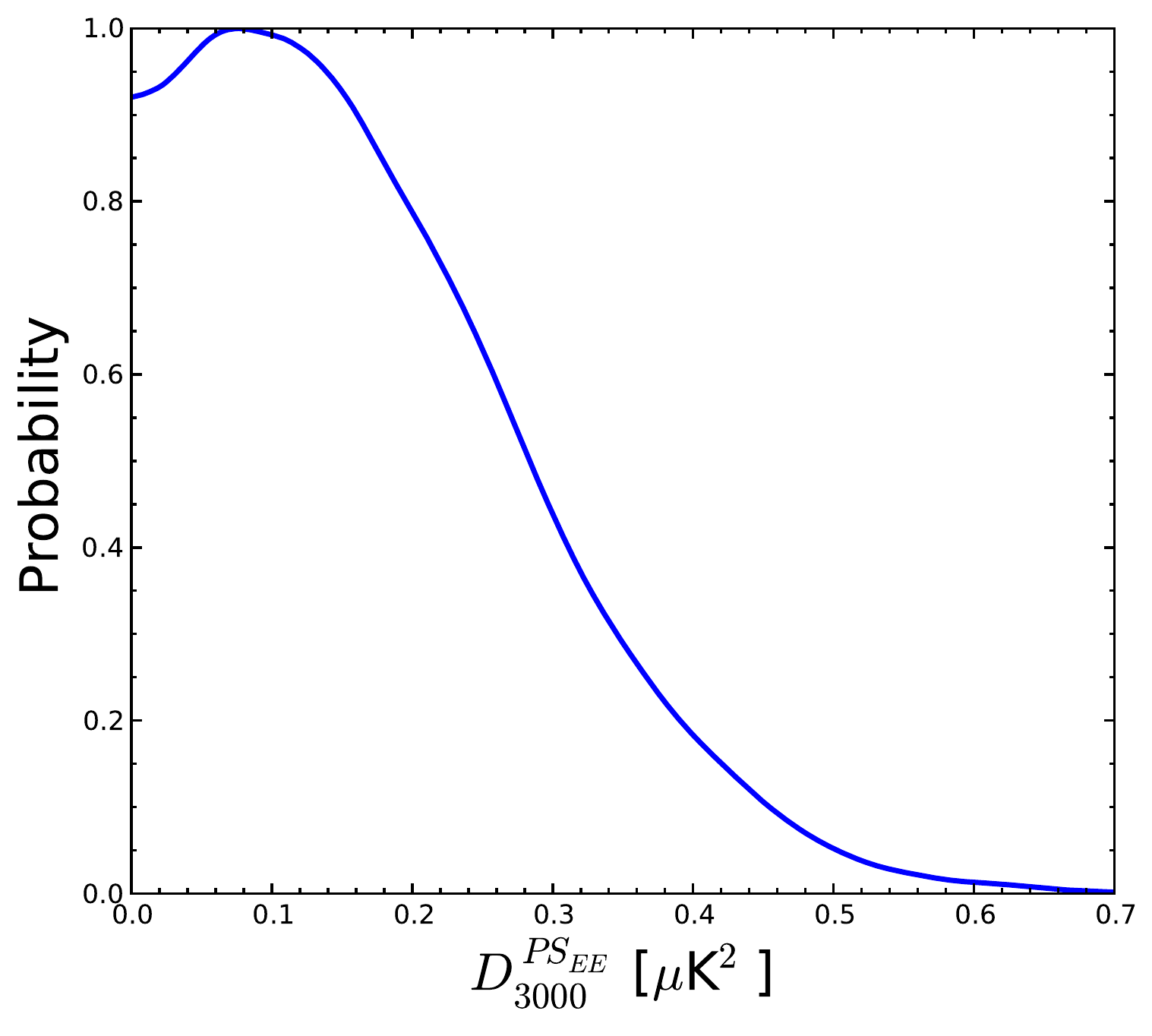}
\end{center}
\caption{1-D marginalized posterior probability for $D^{\mathrm{PS_{EE}}}_{3000}$, the amplitude of residual $EE$ Poisson power in the SPTpol data.}
\label{fig:DpsEE}
\vspace{0.2in}
\end{figure}

The lack of significant high-$\ell$ power in the $EE$ spectrum shown in Figure \ref{fig:spectra_sptpol} indicates 
that, at the level of point-source masking used in this analysis (all sources above 50~mJy in unpolarized flux
masked), polarized point sources do not contribute a significant residual to the $EE$ spectrum.  As a confirmation of this, we find that the nuisance variable that describes this residual, $D^{\mathrm{PS_{EE}}}_{3000}$, has a best-fit amplitude of  $D^{\mathrm{PS_{EE}}}_{3000} =0.07 \pm 0.18 \ \microKsq$.  When we interpret this 
signal as coming from actual sources on the sky, we impose a $D^{\mathrm{PS_{EE}}}_{3000} > 0$ prior; the 
resulting posterior probability distribution for $D^{\mathrm{PS_{EE}}}$ is shown in Figure \ref{fig:DpsEE}.
There is clearly no detection of $D^{\mathrm{PS_{EE}}}_{3000}$ from our data, so we compute the $95\%$
confidence upper limit to this parameter and find $D^{\mathrm{PS_{EE}}}_{3000} < 0.40 \ \microKsq$. 
This corresponds to an upper limit on a constant-in-$\ell$ value of $C^{\mathrm{PS_{EE}}}_{\ell} < 2.8 \times 10^{-7} \ \microKsq$, or $<1.8 \ \microK$-arcmin 
rms fluctuations in the $E$-mode map contributed by Poisson sources.

The recent $EE$ spectrum
measurement from the ACTPol collaboration placed an upper limit of
$D^{\mathrm{PS_{EE}}}_{3000} < 2.4 \ \microKsq$ at  $95\%$ confidence with no sources masked \citep{naess14}. 
The limit reported here 
improves upon this by a factor of six, partially due to the source masking, but 
mostly through higher sensitivity at high $\ell$. (This can be inferred from the fact that
neither experiment has detected the Poisson signal at high significance.)

Using this upper limit as the
amplitude of an $\ell^2$ term, we find that this signal crosses our best-fit $EE$ spectrum at 
$\ell \simeq 3300$.  In a future survey with higher signal-to-noise, this limit could be extended to higher $\ell$ with a more aggressive point source masking. We note that the 50\,mJy threshold used in this work was not limited by source detection; in principle sources could have been masked down to at least 5\,mJy in unpolarized flux, and future experiments could mask even more aggressively. The point-source power in $TT$ is reduced by at least $50\%$ when the source cut is lowered from 50\,mJy to 6\,mJy \citep{mocanu13,george14}; if we assume the $EE$ power is similarly reduced, the resulting $95\%$ upper limit to polarized source power with a 6\,mJy cut would cross our best-fit $EE$ spectrum at $\ell \simeq 3600$.

Under the assumption that the polarization angles of extragalactic sources are randomly distributed
(such that polarized point sources contribute equal $E$-mode and $B$-mode signal),
the anisotropy power contributed by point sources to the $EE$ spectrum is 
equal to the product of the point-source anisotropy power in the $TT$ spectrum and the 
flux-weighted, mean-squared polarization fraction of those sources.
Thus, using previous measurements of the $TT$ point-source anisotropy power, 
our limit on polarized point-source power can be translated into an upper limit on the mean-squared
polarization of sources. 
With a 50 mJy cut, point-source power in the $TT$ spectrum is expected to be 
roughly equally distributed between synchrotron-dominated and dust-dominated sources, with 
$D^{\mathrm{PS_{TT}}}_{3000} \simeq 9 \ \microKsq$ from each component \citep{mocanu13,george14}.
If we assume roughly equal contribution to $EE$ from each type of source, we find a $95\%$ upper limit 
to the mean-squared polarization of sources of $0.021$, or an upper limit to rms polarization
fraction of $14\%$. If we instead assume that the polarization is dominated by the 
synchrotron sources, we find an upper limit to the mean-squared polarization of those sources of 
$0.041$, or an upper limit to rms polarization
fraction of $20\%$. These limits are significantly higher than estimates in the literature of the polarization
fraction of either type of source \citep[e.g.,][]{seiffert07,battye11}; we therefore expect the contamination
from point sources to future $EE$ measurements to be even smaller than the limits considered above.

\section{Conclusion}
\label{sec:conclusion}

We have presented measurements of the $TE$ and $EE$ CMB polarization power spectra using 
data from the first season of observations with SPTpol.
The third through eighth acoustic peaks in the $EE$ spectrum are measured with high signal-to-noise, 
and the measurements of both spectra above $\ell \simeq 1500$ are the most sensitive to date.  We have shown that the SPTpol dataset is consistent with the \LCDM cosmological model preferred by previously published $TT$ spectra from the \planck\ and SPT-SZ instruments.  The inclusion of SPTpol 100d data in cosmological fits marginally improves cosmological parameter 
uncertainties, but due to the small map area and consequently large sample variance, this initial data set has limited cosmological constraining power.
However, the high-$\ell$ sensitivity leads to a significant improvement in the upper limits on the polarized 
point-source power: the constraint on Poisson point-source power in the $EE$ spectrum at $\ell=3000$ is
$D^{\mathrm{PS_{EE}}}_{3000} < 0.40 \ \microKsq$ at $95\%$ confidence, when masking sources with unpolarized flux $> 50\,$mJy. This represents an improvement of a factor of six over previous limits and indicates that
power from uncorrelated point sources will not be a limiting factor to future $EE$ measurements in 
the multipole range $\ell < 3600$, and possibly much higher in $\ell.$

High-fidelity measurements of the $TT$, $TE$, and $EE$ power spectra now span a wide range of scales, from several arcminutes to tens of degrees.  Current small-angular-scale CMB polarization measurements, including the SPTpol data presented here, mainly serve to provide another precision test of the base \LCDM\ model.  More precise measurements of the photon-diffusion-damped region of the polarized power spectra have the potential to place tight constraints on physics that alters the photon diffusion scale, such as the amount of primordial helium, $Y_\mathrm{p}$, and the effective number of relativistic species, $\neff$.  Such measurements will be available in the very near future with data from the completed SPTpol 500~deg$^2$ survey and from other instruments, such as ACTPol, \planck, and POLARBEAR.  
With these upcoming datasets, we will begin to probe the physics revealed by the polarized CMB anisotropy on fine angular scales which will complement previous studies of CMB temperature anisotropy.

The contribution to high-$\ell$ power from residual point-source power and secondary anisotropies like the 
thermal and kinetic SZ effects to the $TE$ and $EE$ spectra are expected to be far smaller than in the $TT$ power spectrum.
This is confirmed for point-source power by the limit presented above, which is a factor of $\sim 50$
below the measured $TT$ point-source power with the same source cut.  Recent forecasts even suggest that the polarization power spectra can constrain the base \LCDM{} cosmological parameters better than the temperature power spectrum, given sufficient sky coverage and sensitivity \citep{galli14}.  The measurements of the high-$\ell$ $TE$ and $EE$ spectra presented here represent an important step towards exploiting the cosmological power of measurements of the polarized damping tails.  

\acknowledgements{
We thank Joshua Schiffrin for help with the mode-mode coupling kernel calculation.  We also thank Wayne Hu and Aurlien Benoit-L\'evy for help implementing the super-sample lensing covariance.  The McGill authors acknowledge funding from the Natural Sciences and Engineering Research Council of Canada, Canadian Institute for Advanced Research, and Canada Research Chairs program.
The South Pole Telescope program is supported by the National Science Foundation through grant PLR-1248097. Partial support is also provided by the NSF Physics Frontier Center grant PHY-0114422 to the Kavli Institute of Cosmological Physics at the University of Chicago, the Kavli Foundation, and the Gordon and Betty Moore Foundation through Grant GBMF\#947 to the University of Chicago.  
JWH is supported by the National Science Foundation under Award No. AST-1402161.
BB is supported by the Fermi Research Alliance, LLC under Contract No. De-AC02-07CH11359 with the U.S. Department of Energy.  
The CU Boulder group acknowledges support from NSF AST-0956135.  
This work is also supported by the U.S. Department of Energy.  
Work at Argonne National Lab is supported by UChicago Argonne, LLC, Operator of Argonne National Laboratory (Argonne). 
Argonne, a U.S. Department of Energy Office of Science Laboratory, is operated under Contract No. DE-AC02-06CH11357. 
We also acknowledge support from the Argonne Center for Nanoscale Materials.  
The data analysis pipeline uses the scientific python stack \citep{hunter07, jones01, vanDerWalt11} and the HDF5 file format \citep{hdf5}.
}

\bibliography{spt}

\appendix
\label{appendix:mode_mixing}

In this appendix, we calculate the mode-mode coupling kernel for the $EE$ and $TE$ power spectra.
This calculation closely follows the flat-sky calculation in Appendix A of \cite{hivon02},
but includes polarization as well as temperature

To remove point sources and taper the edges of our finite-sky coverage, 
we multiply our $T$, $Q$, and $U$ maps by the mask in Figure \ref{fig:mask}.
This multiplication in 2-D real space is a convolution in Fourier space.
This has the effect of mixing 2-D Fourier modes of different $\pmb{\ell}$.
The pseudo power spectrum $\tilde{C}_{\ell}$ is the azimuthal average 
of this altered 2-D spectrum.
The mode-mode coupling kernel is an analytical expression of how the mask 
alters the underlying Gaussian spectrum.

For temperature, the mode-mixing kernel accounts for this convolution and azimuthal averaging.
For polarization, the mask is applied to the $Q$ and $U$ maps, 
but there is an additional step to rotate into $E$ and $B$ before azimuthally averaging.
This rotation changes the form of the coupling kernel and also introduces $E/B$ mixing.
The effect that this has on the $TE$ and $EE$ power spectra is calculated here,
under the assumption that the underlying $BB$ power is negligible.

We write a field $X(\mb{r})$ on the plane with Fourier conjugate $X(\mb{k})$ as
\begin{equation}
X(\mb{r})   = \int{d\mb{k} \,\, X(\mb{k}) e^{2i\pi \mb{k}\cdot\mb{r}} }
 \leftrightarrow 
X(\mb{k}) = \int{d\mb{r}\, X(\mb{r}) e^{-2i\pi \mb{k}\cdot\mb{r}} }\, ,
\end{equation}
for $X \in [T, Q, U]$ and $\mb{k} \propto \pmb{\ell}$.
We transform $Q$,$U$ to $E$,$B$ as shown in Eq. \ref{eqn:def_E} of the main text, using the conventions defined in \cite{zaldarriaga01}, where
a particular 2-D Fourier component of $Q$ and $U$ is a simple linear combination (a rotation) of $E$ and $B$ for that same $\mb{k}$:
\begin{eqnarray}
Q(\mb{k_1}) &= [E(\mb{k_1}) \cos(2\phi_{k_1}) - \,B(\mb{k_1})\sin(2\phi_{k_1})] &= [E_1 C_1 - \,B_1 S_1]\\\nonumber
U(\mb{k_1}) &= [E(\mb{k_1}) \sin(2\phi_{k_1}) + \,B(\mb{k_1})\cos(2\phi_{k_1})] &= [E_1 S_1 + \,B_1 C_1] \ ,
\end{eqnarray}
where we have used the following abbreviations to indicate the dependence on the first of many $\mb{k}$'s: \\
$C_1 \equiv \cos(2\phi_{k_1}), S_1 \equiv \sin(2\phi_{k_1}) \hbox{ and } X_1 \equiv\, X(\mb{k_1})$.
Solving these equations for $E$, $B$ (reversing the rotation), we find 
\begin{eqnarray}
E_1 &=& [Q_1 C_1 + \,U_1 S_1] \\ \nonumber
B_1 &=& [U_1 C_1 - \,Q_1 S_1] \ .
\end{eqnarray}

The altered ``pseudo-'' Fourier components on a plane that has been masked or weighted by $W(\mb{r})$ can be written as a convolution
\begin{eqnarray}
  \tilde{X}(\mb{k_1}) 
  &=& \int{d\mb{r}\, X(\mb{r}) W(\mb{r}) e^{-2i\pi \mb{k_1}\cdot\mb{r}} } \\\nonumber
  &=& \int{d\mathbf{k_2}\, X(\mathbf{k_2}) W(\mathbf{k_1}-\mathbf{k_2}) } \\\nonumber
  &=& \int{d\mathbf{k_2}\, X(\mathbf{k_2}) K_{\mathbf{k_2,k_1}} } .
\end{eqnarray} 
Here, $K_{\mathbf{k_2,k_1}} \equiv \int{d\mathbf{k'}\, W(\mathbf{k'})\delta(-\mathbf{k_1}+\mathbf{k_2}+\mathbf{k'})}$ is the same as the scalar case, eqn (A7, A8) from \cite{hivon02}.

Now calculate the pseudo power spectrum $\langle\,_{EE}\tilde{C}_{k_1}\rangle$ by azimuthally averaging. For each Forier magnitude $k_1$,
\begin{eqnarray}
\label{eqn:ps_qu}
  \langle\,_{EE}\tilde{C}_{k_1}\rangle & \equiv &
  \frac{1}{2\pi} \int{ d\phi_1\, \langle\,\tilde{E}_1 \,\tilde{E}_1^* \rangle } \\\nonumber
  &=& \frac{1}{2\pi} \int{ d\phi_1\, \left[
    \langle\,\tilde{Q}_1 \,\tilde{Q}_1^* \rangle C_1^2 + 
    \langle\,\tilde{U}_1 \,\tilde{U}_1^* \rangle S_1^2 + 
    \langle\,\tilde{Q}_1 \,\tilde{U}_1^* \rangle C_1 S_1 + 
    \langle\,\tilde{Q}_1^* \,\tilde{U}_1 \rangle C_1 S_1
\right] } \ .
\end{eqnarray}

Consider the first term and write the pseudo-$\tilde{Q}$ in terms of a convolution of the unmasked-$Q$, 

\begin{equation} 
   \frac{1}{2\pi} \int{ d\phi_1\,  \langle\,\tilde{Q}_1 \,\tilde{Q}_1^* \rangle C_1^2 } 
=  \frac{1}{2\pi} \int{ d\phi_1\, \int{\int{d\mathbf{k_2}\, d\mathbf{k_3}\, 
                                       \langle \,Q_2 \,Q_3^* \rangle  
                                       K_{\mathbf{k_2,k_1}} 
                                       K_{\mathbf{k_3,k_1}}^* C_1^2 }}} .
\end{equation}
Now write $Q$ in terms of $E$ and $B$.
Assuming that $E$ and $B$ are homogeneous, isotropic, Gaussian-distributed fields,
we can use $\langle \,E_{k_2} \,E_{k_3}^* \rangle = \langle \,_{EE}C_{k_2} \rangle\delta(\mathbf{k_2 - k_3} )$ to write this term as:
\begin{eqnarray} 
&=&  \frac{1}{2\pi} \int{ d\phi_1\, \int{\int{d\mathbf{k_2}\, d\mathbf{k_3}\,  \left[
   \langle\,_{EE}C_{k_2}\rangle C_2 C_3
  + \langle\,_{BB}C_{k_2}\rangle S_2 S_3 \right.}}} \\\nonumber
&& \qquad \qquad \qquad \qquad \qquad \qquad
\left. - \langle\,_{EB}C_{k_2}\rangle C_2 S_3
  - \langle\,_{EB}C_{k_2}\rangle S_2 C_3  \right] \delta(\mathbf{k_2-k_3})   K_{\mathbf{k_2,k_1}} K_{\mathbf{k_3,k_1}}^* C_1^2 .
\end{eqnarray}
Integrate over $\mathbf{k_3}$ using $\delta(\mathbf{k_2-k_3})$:
\begin{equation} 
=  \frac{1}{2\pi} \int{ d\phi_1\, \int{d\mathbf{k_2}\, \left[
  + \langle\,_{EE}C_{k_2}\rangle C_2^2
  + \langle\,_{BB}C_{k_2}\rangle S_2^2
  -2\langle\,_{EB}C_{k_2}\rangle C_2 S_2
 \right] C_1^2
 \,\left| K_{\mathbf{k_2,k_1}} \right|^2 }} .
\end{equation}
Working out all four terms in Eq. \ref{eqn:ps_qu} yields 
\begin{eqnarray}
\label{eqn:ps_mm_def}
\langle\,_{EE}\tilde{C}_{k_1}\rangle & =\,
  \frac{1}{2\pi} \int{ d\phi_1\, \int{d\mathbf{k_2}\, }} [ &
   \langle_{EE}C_{k_2}\rangle (C_2 C_1 + S_2 S_1)^2 + \\\nonumber
&& \langle_{BB}C_{k_2}\rangle (S_2 C_1 - C_2 S_1)^2 + \\\nonumber
&& \langle_{EB}C_{k_2}\rangle 2(S_1 C_1(C_2^2-S_2^2) - S_2 C_2(C_1^2 - S_1^2)) ]\,
\left| K_{\mathbf{k_2,k_1}} \right|^2 \\
 &= \int{dk_2\,k_2} [ & \!\!\!\!\!\!\!\!\! \mathbf{M}^{EE}_{k_1\,k_2}\langle_{EE}C_{k_2}\rangle + \mathbf{M}^{BB}_{k_1\,k_2}\langle_{BB}C_{k_2}\rangle + \mathbf{M}^{EB}_{k_1\,k_2}\langle_{EB}C_{k_2}\rangle ] \ .
\end{eqnarray}
The coupling kernel $\mathbf{M}^{XY}_{k_1\,k_2}$, where $X,Y \in \{E, B\}$, depends only on the magnitude of $\mathbf{k_1}$ and $\mathbf{k_2}$ and is given by
\begin{equation}
\label{eqn:mm_xy}
  \mathbf{M}^{XY}_{k_1\,k_2} \equiv
  \int{dk'\,k'\, \mathcal{W}(k') \int{ \!\! \int{d\phi_1\,d\phi_2\, \ A^{XY}_{1,2}\, \delta(-\mathbf{k_1}+\mathbf{k_2}+\mathbf{k'}) }}} ,
\end{equation}
where $2\pi \mathcal{W}(k) = \int{d\phi \, W(\mathbf{k})W(\mathbf{k})^*}$ is the azimuthally-integrated power spectrum of the mask.
Here, the coefficients $A^{XY}_{1,2}$ correspond to the trigonometric coefficients in Eq. \ref{eqn:ps_mm_def}.
Specifically, $A^{EE}_{1,2} = (C_2 C_1 + S_2 S_1)^2$.
In this paper, we assume that $\langle_{EE}C_{k_2}\rangle$ is much bigger than either $\langle_{EB}C_{k_2}\rangle$ or $\langle_{BB}C_{k_2}\rangle$, and consequently ignore the contributions to $\langle\,_{EE}\tilde{C}_{k_1}\rangle$ from those terms in Eq. \ref{eqn:ps_mm_def}.

We want to evaluate the integrals over $\phi_1$ and $\phi_2$ in Eq. \ref{eqn:mm_xy}; to accomplish this we need to evaluate expressions of the form $\iint\limits d\phi_1\, d\phi_2\, \delta(-\mathbf{k_1}+\mathbf{k_2}+\mathbf{k'}) G(\phi_1, \phi_2)$.
Make the variable substitution $\mb{k_4} \equiv ( \mb{k_2 + k'})$ and use the relations $\delta(\mathbf{r} - \mathbf{r'}) = \frac{1}{r}\delta(r-r')\delta(\phi-\phi')$ and 
\[\delta\left(g(\phi_2)\right) = \sum_i \frac{\delta\left(\phi_2-\phi_2^{(i)}\right)}{\left| g'(\phi_2^{(i)}) \right|},\]
where $\phi_2^{(i)}$ are the $i$ roots of $g(\phi_2^{(i)}) = 0$ between 0 and $2\pi$.
If $G(\phi_1, \phi_2)=1$, as in the $\mb{M}^{TT}$ case where $A^{TT}_{1,2} = 1$, then
\begin{eqnarray}
  \iint\limits d\phi_1\, d\phi_2\, \delta(-\mathbf{k_1}+\mathbf{k_2}+\mathbf{k'}) &=
  \frac{1}{k_1} \int{d\phi_2 \delta \left(g(\phi_2)\right) } \\\nonumber
  &= \frac{1}{k_1}\sum_i \frac{1}{\left| g'(\phi_2^{(i)}) \right|} \\\nonumber
  &= 2\pi J(k_1,k_2,k'),
\end{eqnarray}
where $g(\phi_2) \equiv k_1 - \sqrt{k_2^2 + k'^2 + 2k_2 k' \cos(\phi' - \phi_2)}$, 
and $J(k_1,k_2,k') \equiv (\frac{2}{\pi}) \left[2k_1^2k_2^2 + 2k_1^2k'^2 + 2k_2^2\,k'^2 - k_1^4 - k_2^4 - k'^4\right]^{-1/2}$
for $|k_2-k'|<k_1<k_2+k'$, and $J=0$ otherwise.  
This is the result derived in Eq. (A10) from \cite{hivon02}.

In the calculation of $\mb{M}^{EE}$, $G(\phi_1, \phi_2)=(C_2 C_1 + S_2 S_1)^2 = \cos^2[2(\phi_{2} - \phi_{1})]$.
We derive an expression for $G(\phi_1, \phi_2)$ as follows: we define $\mb{k_4} \equiv ( \mb{k_2 + k'})$ and without loss of generality set $\phi'=0$; from this we derive the useful relation 
\[\cos(\phi_4) = \frac{k_2\cos(\phi_2) + k'}{\sqrt{k_2^2 + k'^2 + 2k_2k'\cos(\phi_2)} } .\]
We take advantage of the fact that $\cos(\phi_4)$ is only a function of $\phi_2$ in order to write
\begin{flalign}
  G(\phi_1, \phi_2) 
  &= \cos^2[2(\phi_{2} - \phi_{4})] \\\nonumber
  &= \left(2\cos^2(\phi_2-\phi_4) - 1 \right)^2 \\\nonumber
  &= \left[2\left[\cos(\phi_4)\cos(\phi_2) + \sqrt{(1-\cos^2(\phi_4)) \cdot (1-\cos^2(\phi_2))} \right]^2 - 1 \right]^2 \\\nonumber
  &\equiv G(\cos(\phi_2)).
\end{flalign}
Using this form we can write
\begin{flalign}
  \iint\limits d\phi_1\, d\phi_2\, \delta(-\mathbf{k_1}+\mathbf{k_2}+\mathbf{k'}) G(\phi_1, \phi_2)
  &= \frac{1}{k_1}\int{d\phi_2\,\delta(g(\phi_2)) \,G(\cos(\phi_2))} \\\nonumber
  &= \frac{1}{k_1} \sum_{(+,-)} \frac{G(\cos(\phi_2^{(+,-)}))}{\left| g'(\phi_2^{(+,-)})\right|}.
\end{flalign}
Here, $g(\phi_2) = k_1 - \sqrt{k_2^2 + k'^2 + 2k_2 k' \cos(\phi' - \phi_2)} = k_1 - \left|\mathbf{k_2} + \mathbf{k'}\right|$.\\
$g(\phi_2)$ has the same two roots, $\phi_2^{(i)}=\phi_2^{(+,-)}$, as in the TT calculation.
Specifically,
\begin{equation}
  \phi_2^i = \phi_2^{(+,-)} = \pm \cos^{-1}\left(\frac{k_1^2 - (k_2^2+k'^2)}{2k_2\,k'} \right).
\end{equation}
We find $\cos(\phi_2^+) = \cos(\phi_2^-)$, and evaluate the expression for $G(\phi_1, \phi_2)$ at these roots:
\begin{equation}
  G(\cos(\phi_2^{(+)})) = \left[ 1 - \frac{(k_1^2+k_2^2-k'^2)^2}{2 k_1^2 k_2^2} \right]^2 \, .
\end{equation}
Thus the $EE$ coupling kernel is
\begin{equation}
  \mathbf{M}^{EE}_{k_1\,k_2} = \int{dk'\,k'\, \mathcal{W}(k') } 2\pi J(k_1,k_2,k') \cdot 
  \left[ 1 - \frac{(k_1^2+k_2^2-k'^2)^2}{2 k_1^2 k_2^2} \right]^2.
\end{equation}

The calculation of the mode-coupling kernel for $\langle\,_{TE}\tilde{C}_{k_1}\rangle$ proceeds similarly.
We summarize this calculation as follows:
\begin{equation} \langle\,_{TE}\tilde{C}_{k_1}\rangle = \int{dk_2\,k_2 
  \left[\mathbf{M}^{TE}_{k_1\,k_2}\langle_{TE}C_{k_2}\rangle + \mathbf{M}^{TB}_{k_1\,k_2}\langle_{TB}C_{k_2}\rangle \right]},\end{equation}
where
\begin{equation}
  \mathbf{M}^{TE}_{k_1\,k_2} \equiv
  \int{dk_3\,k_3\, \mathcal{W}(k_3) \int{\int{d\phi_1\,d\phi_2\, \cos(2(\phi_{k_2} - \phi_{k_1}))\, \delta(-\mathbf{k_1}+\mathbf{k_2}+\mathbf{k_3}) }}} \, .
\end{equation}

We assume that $\langle\,_{TE}\tilde{C}_{k_1}\rangle$ is much larger than $\langle\,_{TB}\tilde{C}_{k_1}\rangle$, and thus only consider $\mathbf{M}^{TE}_{k_1\,k_2}$.
The final expression for the $TE$ coupling kernel is:
\begin{equation}
  \mathbf{M}^{TE}_{k_1\,k_2} = \int{dk'\,k'\, \mathcal{W}(k') } 2\pi J(k_1,k_2,k') * 
    \left[ 1 - \frac{(k_1^2+k_2^2-k'^2)^2}{2 k_1^2 k_2^2} \right] \, .
\end{equation}

\end{document}

%% file: ee2013_authorlist_v3.tex
\def\AAUChicago{1}
\def\KICPChicago{2}
\def\Caltech{3}
\def\ColoradoAPS{4}
\def\Cardiff{5}
\def\UChicago{6}
\def\NIST{7}
\def\McGill{8}
\def\ArgonneHEP{9}
\def\FNAL{10}
\def\PhysicsUChicago{11}
\def\EFIChicago{12}
\def\UKZN{13}
\def\SLAC{14}
\def\CIFAR{15}
\def\Berkeley{16}
\def\ColoradoPhys{17}
\def\Stanford{18}
\def\KIPAC{19}
\def\Davis{20}
\def\LBNL{21}
\def\Michigan{22}
\def\CaseWestern{23}
\def\ArgonneMSD{24}
\def\Minnesota{25}
\def\Melbourne{26}
\def\ArtInstChicago{27}
\def\ThreeSpeedLogic{28}
\def\CfA{29}
\def\Dunlap{30}
\def\UToronto{31}
\def\BCCP{32}
\def\illast{33}
\def\illphy{34}


 \author{
  A.T.~Crites\altaffilmark{\AAUChicago,\KICPChicago,\Caltech},
  J.~W.~Henning\altaffilmark{\KICPChicago,\ColoradoAPS},
  P.~A.~R.~Ade\altaffilmark{\Cardiff},
  K.~A.~Aird\altaffilmark{\UChicago},
  J.~E.~Austermann\altaffilmark{\ColoradoAPS},
  J.~A.~Beall\altaffilmark{\NIST} ,
  A.~N.~Bender\altaffilmark{\McGill,\ArgonneHEP},
  B.~A.~Benson\altaffilmark{\AAUChicago,\KICPChicago,\FNAL},
  L.~E.~Bleem\altaffilmark{\KICPChicago,\ArgonneHEP,\PhysicsUChicago},
  J.~E.~Carlstrom\altaffilmark{\AAUChicago,\KICPChicago,\ArgonneHEP,\PhysicsUChicago,\EFIChicago},
  C.~L.~Chang\altaffilmark{\AAUChicago,\KICPChicago,\ArgonneHEP},
  H.~C.~Chiang\altaffilmark{\UKZN},
  H-M.~Cho\altaffilmark{\SLAC},
  R.~Citron\altaffilmark{\KICPChicago},
  T.~M.~Crawford\altaffilmark{\AAUChicago,\KICPChicago},
  T.~de~Haan\altaffilmark{\McGill},
  M.~A.~Dobbs\altaffilmark{\McGill,\CIFAR},
  W.~Everett\altaffilmark{\ColoradoAPS},
  J.~Gallicchio\altaffilmark{\KICPChicago},
  J.~Gao\altaffilmark{\NIST},
  E.~M.~George\altaffilmark{\Berkeley},
  A.~Gilbert\altaffilmark{\McGill},
  N.~W.~Halverson\altaffilmark{\ColoradoAPS,\ColoradoPhys},
  D.~Hanson\altaffilmark{\McGill},
  N.~Harrington\altaffilmark{\Berkeley},
  G.~C.~Hilton\altaffilmark{\NIST},
  G.~P.~Holder\altaffilmark{\McGill},
  W.~L.~Holzapfel\altaffilmark{\Berkeley},
  S.~Hoover\altaffilmark{\KICPChicago,\PhysicsUChicago},
  Z.~Hou\altaffilmark{\KICPChicago},
  J.~D.~Hrubes\altaffilmark{\UChicago},
  N.~Huang\altaffilmark{\Berkeley},
  J.~Hubmayr\altaffilmark{\NIST},
  K.~D.~Irwin\altaffilmark{\SLAC,\Stanford},
  R.~Keisler\altaffilmark{\Stanford,\KIPAC},
  L.~Knox\altaffilmark{\Davis},
  A.~T.~Lee\altaffilmark{\Berkeley,\LBNL},
  E.~M.~Leitch\altaffilmark{\AAUChicago,\KICPChicago},
  D.~Li\altaffilmark{\NIST,\SLAC},
  C.~Liang\altaffilmark{\UChicago},       
  D.~Luong-Van\altaffilmark{\UChicago},
  J.~J.~McMahon\altaffilmark{\Michigan},
  J.~Mehl\altaffilmark{\KICPChicago,\ArgonneHEP},
  S.~S.~Meyer\altaffilmark{\AAUChicago,\KICPChicago,\PhysicsUChicago,\EFIChicago},
  L.~Mocanu\altaffilmark{\AAUChicago,\KICPChicago},
  T.~E.~Montroy\altaffilmark{\CaseWestern},
  T.~Natoli\altaffilmark{\KICPChicago,\PhysicsUChicago},
  J.~P.~Nibarger\altaffilmark{\NIST},
  V.~Novosad\altaffilmark{\ArgonneMSD},
  S.~Padin\altaffilmark{\AAUChicago,\KICPChicago,\Caltech},
  C.~Pryke\altaffilmark{\Minnesota},
  C.~L.~Reichardt\altaffilmark{\Berkeley,\Melbourne},
  J.~E.~Ruhl\altaffilmark{\CaseWestern},
  B.~R.~Saliwanchik\altaffilmark{\CaseWestern},
  J.T.~Sayre\altaffilmark{\CaseWestern},
  K.~K.~Schaffer\altaffilmark{\KICPChicago,\EFIChicago,\ArtInstChicago},
  G.~Smecher\altaffilmark{\McGill,\ThreeSpeedLogic},
  A.~A.~Stark\altaffilmark{\CfA},
  K.T.~Story\altaffilmark{\KICPChicago},
  C.~Tucker\altaffilmark{\Cardiff},
  K.~Vanderlinde\altaffilmark{\Dunlap,\UToronto},
  J.~D.~Vieira\altaffilmark{\Caltech,\illast,\illphy},
  G.~Wang\altaffilmark{\ArgonneHEP},
  N.~Whitehorn\altaffilmark{\Berkeley},
  V.~Yefremenko\altaffilmark{\ArgonneHEP},
  and 
  O.~Zahn\altaffilmark{\BCCP}
 }

\altaffiltext{\AAUChicago}{Department of Astronomy and Astrophysics, University of Chicago, 5640 South Ellis Avenue, Chicago, IL, USA 60637}
\altaffiltext{\KICPChicago}{Kavli Institute for Cosmological Physics, University of Chicago, 5640 South Ellis Avenue, Chicago, IL, USA 60637}
\altaffiltext{\Caltech}{California Institute of Technology, MS 249-17, 1216 E. California Blvd., Pasadena, CA, USA 91125}
\altaffiltext{\ColoradoAPS}{Department of Astrophysical and Planetary Sciences, University of Colorado, Boulder, CO, USA 80309}
\altaffiltext{\Cardiff}{Cardiff University, Cardiff CF10 3XQ, United Kingdom}
\altaffiltext{\UChicago}{University of Chicago, 5640 South Ellis Avenue, Chicago, IL, USA 60637}
\altaffiltext{\NIST}{NIST Quantum Devices Group, 325 Broadway Mailcode 817.03, Boulder, CO, USA 80305}
\altaffiltext{\McGill}{Department of Physics, McGill University, 3600 Rue University, Montreal, Quebec H3A 2T8, Canada}
\altaffiltext{\ArgonneHEP}{High Energy Physics Division, Argonne National Laboratory,9700 S. Cass Avenue, Argonne, IL, USA 60439}
\altaffiltext{\FNAL}{Fermi National Accelerator Laboratory, MS209, P.O. Box 500, Batavia, IL 60510}
\altaffiltext{\PhysicsUChicago}{Department of Physics, University of Chicago, 5640 South Ellis Avenue, Chicago, IL, USA 60637}
\altaffiltext{\EFIChicago}{Enrico Fermi Institute, University of Chicago, 5640 South Ellis Avenue, Chicago, IL, USA 60637}
\altaffiltext{\UKZN}{School of Mathematics, Statistics \& Computer Science, University of KwaZulu-Natal, Durban, South Africa}
\altaffiltext{\SLAC}{SLAC National Accelerator Laboratory, 2575 Sand Hill Road, Menlo Park, CA 94025}
\altaffiltext{\CIFAR}{Canadian Institute for Advanced Research, CIFAR Program in Cosmology and Gravity, Toronto, ON, M5G 1Z8, Canada}
\altaffiltext{\Berkeley}{Department of Physics, University of California, Berkeley, CA, USA 94720}
\altaffiltext{\ColoradoPhys}{Department of Physics, University of Colorado, Boulder, CO, USA 80309}
\altaffiltext{\Stanford}{Dept. of Physics, Stanford University, 382 Via Pueblo Mall, Stanford, CA 94305}
\altaffiltext{\KIPAC}{Kavli Institute for Particle Astrophysics and Cosmology, Stanford University, 452 Lomita Mall, Stanford, CA 94305}
\altaffiltext{\Davis}{Department of Physics, University of California, One Shields Avenue, Davis, CA, USA 95616}
\altaffiltext{\LBNL}{Physics Division, Lawrence Berkeley National Laboratory, Berkeley, CA, USA 94720}
\altaffiltext{\Michigan}{Department of Physics, University of Michigan, 450 Church Street, Ann  Arbor, MI, USA 48109}
\altaffiltext{\CaseWestern}{Physics Department, Center for Education and Research in Cosmology and Astrophysics, Case Western Reserve University,Cleveland, OH, USA 44106}
\altaffiltext{\ArgonneMSD}{Materials Sciences Division, Argonne National Laboratory,9700 S. Cass Avenue, Argonne, IL, USA 60439}
\altaffiltext{\Minnesota}{School of Physics and Astronomy, University of Minnesota, 116 Church Street S.E. Minneapolis, MN, USA 55455}
\altaffiltext{\Melbourne}{School of Physics, University of Melbourne, Parkville, VIC 3010, Australia}
\altaffiltext{\ArtInstChicago}{Liberal Arts Department, School of the Art Institute of Chicago, 112 S Michigan Ave, Chicago, IL, USA 60603}
\altaffiltext{\ThreeSpeedLogic}{Three-Speed Logic, Inc., Vancouver, B.C., V6A 2J8, Canada}
\altaffiltext{\CfA}{Harvard-Smithsonian Center for Astrophysics, 60 Garden Street, Cambridge, MA, USA 02138}
\altaffiltext{\Dunlap}{Dunlap Institute for Astronomy \& Astrophysics, University of Toronto, 50 St George St, Toronto, ON, M5S 3H4, Canada}
\altaffiltext{\UToronto}{Department of Astronomy \& Astrophysics, University of Toronto, 50 St George St, Toronto, ON, M5S 3H4, Canada}
\altaffiltext{\BCCP}{Berkeley Center for Cosmological Physics, Department of Physics, University of California, and Lawrence Berkeley National Laboratory, Berkeley, CA, USA 94720}
\altaffiltext{\illast}{Astronomy Department, University of Illinois at Urbana-Champaign, 1002 W.\ Green Street, Urbana, IL 61801, USA}
\altaffiltext{\illphy}{Department of Physics, University of Illinois Urbana-Champaign, 1110 W.\ Green Street, Urbana, IL 61801, USA}

\email{acrites@caltech.edu}